\documentclass[aip,adv,reprint,floatfix]{revtex4-2}

\usepackage{syntonly}

\usepackage[utf8]{inputenc}

\usepackage[english]{babel}

\usepackage{xcolor}
\definecolor{lightblue}{RGB}{166,206,227}
\definecolor{darkblue}{RGB}{31,120,180}
\definecolor{lightgreen}{RGB}{178,223,138}
\definecolor{darkgreen}{RGB}{51,160,44}
\definecolor{teal}{RGB}{27,158,119}
\definecolor{orange}{RGB}{217,95,2}
\definecolor{purple}{RGB}{117,112,179}
\definecolor{magenta}{RGB}{231,41,138}

\usepackage{booktabs} 
\usepackage{siunitx} 
\usepackage{amsmath,amssymb,commath}

\usepackage{graphicx}

\usepackage[color = orange!30!white]{todonotes} 
\presetkeys{todonotes}{inline}{}

\newcommand{\laplacian}{\mathop{{}\bigtriangleup}\nolimits} 

\newcommand{\Gprime}{\ensuremath{G^{\prime}}} 
\newcommand{\Gdoubleprime}{\ensuremath{G^{\prime\prime}}} 

\newcommand*{\mmnote}[1]{{\color{darkblue} #1}}

\begin{document}
\preprint{One-disc paper} 

\title{Piezo-Electric Shear Rheometry: Further developments in experimental implementation and data extraction}

\author{Mathias Mikkelsen}
\email{mathiasm@ruc.dk}
\affiliation{Glass and Time, IMFUFA, Department of Science and Environment, Roskilde University, Roskilde, Denmark}
\affiliation{Research and Development, Continental Reifen Deutschland GmbH, Hannover, Germany}

\author{Kira L. Eliasen}
\affiliation{Glass and Time, IMFUFA, Department of Science and Environment, Roskilde University, Roskilde, Denmark}

\author{Niclas Lindemann}
\affiliation{Research and Development, Continental Reifen Deutschland GmbH, Hannover, Germany}
\affiliation{Institut für Physikalische Chemie und Elektrochemie, Leibniz Universität Hannover, Hannover, Germany}

\author{Kevin Moch}
\affiliation{Fakultät Physik, Technische Universität Dortmund, 44221  Dortmund, Germany}

\author{Roland Böhmer}
\affiliation{Fakultät Physik, Technische Universität Dortmund, 44221  Dortmund, Germany}

\author{Hossein Ali Karimi-Varzaneh}
\affiliation{Research and Development, Continental Reifen Deutschland GmbH, Hannover, Germany}

\author{Jorge Lacayo-Pineda}
\affiliation{Research and Development, Continental Reifen Deutschland GmbH, Hannover, Germany}
\affiliation{Institut für Anorganische Chemie, Leibniz Universität Hannover, Hannover, Germany}

\author{Bo Jakobsen}
\affiliation{Glass and Time, IMFUFA, Department of Science and Environment, Roskilde University, Roskilde, Denmark}

\author{Kristine Niss}
\affiliation{Glass and Time, IMFUFA, Department of Science and Environment, Roskilde University, Roskilde, Denmark}

\author{Tage Christensen}
\affiliation{Glass and Time, IMFUFA, Department of Science and Environment, Roskilde University, Roskilde, Denmark}

\author{Tina Hecksher}
\email{tihe@ruc.dk}
\affiliation{Glass and Time, IMFUFA, Department of Science and Environment, Roskilde University, Roskilde, Denmark}

\begin{abstract}
	The Piezo-electric Shear Gauge (PSG) [Christensen \& Olsen, Rev. Sci. Instrum. \textbf{66}, 5019, 1995] is a rheometric technique developed to measure the complex shear modulus of viscous liquids near their glass transition temperature.
	We report recent advances to the PSG technique:
	1) The data extraction procedure is optimized which extends the upper limit of the frequency range of the method to between \SIrange[range-phrase = { and }]{50}{70}{\kilo\hertz}.
	2) The measuring cell is simplified to use only one piezo-electric ceramic disc instead of three.
	We present an implementation of this design intended for liquid samples.
	Data obtained with this design revealed that a soft extra spacer is necessary to allow for thermal contraction of the sample in the axial direction.
	Model calculations show that flow in the radial direction is hindered by the confined geometry of the cell when the liquid becomes viscous upon cooling. The method is especially well-suited for -- but not limited to -- glassy materials.
\end{abstract}

\maketitle

\section{Introduction}
\label{sec:introduction}
Mechanical properties, e.g. stiffness and viscosity, of complex materials are important for many applications \cite{ashby_metallic_2006,fischer_rheology_2011} as well as for the fundamental understanding of matter \cite{angell_perspective_1988,mauro_viscosity_2009}.
A number of rheometric techniques which map these properties as a function of frequency, temperature, and in some cases stress or strain amplitude, are in use in research labs and industry \cite{hou_instrument_2005,schroyen_bulk_2020}.
These techniques can roughly be divided into three categories: Quasi-static methods, resonant methods and (sound) propagation methods.

\emph{Quasi-static methods} are techniques that deform the entire sample. These techniques include most standard commercial rheometers, but also sliding plate rheometers using piezo ceramic discs, e.g. Ref.~\onlinecite{athanasiou_high-frequency_2019}.
This entails that the deformation rate must be relatively slow and the sample size small compared to the sound velocity wavelength. 
The upper frequency limit of standard rheometers is determined by the resonance frequency of the driving device (in most cases a rotating shaft), which typically lies around \SI{100}{\hertz}.
The limitation in frequency range leads to the need for construction of master curves if one wants to obtain the entire relaxation function.
This procedure assumes time-temperature superposition (TTS), i.e.~that the shape of the relaxation is conserved, so that each measured temperature gives a different part of the full function that can be shifted with respect to each other to form the master curve.

Standard rheometers are sensitive to low moduli/viscosities, but less optimal for harder samples due to the instrument compliance being comparable to the sample compliance \cite{schroter_dynamic_2006,laukkanen_small-diameter_2017}, i.e.~instead of deforming the sample the instrument itself deforms. Corrections are routinely applied to account for this effect, but the method remains more accurate for soft materials.

\emph{Resonant methods} function by monitoring the resonance of the measuring device while it shifts due to the effect of the visco-elastic sample compared to a freely oscillating device.
Traditional examples of implementations include torsional resonators and cantilever devices \cite{ferry_viscoelastic_1980}.
Recent developments for resonant methods go in the direction of micro-rheology, e.g. using atomic force microscopy\cite{sader_frequency_1998,ahmed_measurement_2001} and MEMS based devices\cite{christopher_development_2010, mather_liquid_2012}.
These techniques work mainly for low-viscosity (the \SI{}{\pascal\second} range) liquids.
Resonance methods typically operate at discrete frequencies in the \si{\kilo\hertz} region and are rather precise.

\emph{Propagation methods} generate deformations that travel through the sample and get detected after some propagation length.
Delay time and wave amplitude changes are then recorded and translated into sound velocity and attenuation coefficient.
These methods include classical ultrasonic techniques in the \SIrange{1}{10}{\mega\hertz} region \cite{mcskimin_ultrasonic_1964} and laser techniques such as Impulsive Stimulated Scattering \cite{yan_impulsive_1987,pick_frequency_2003} in the \SIrange[scientific-notation = engineering]{100}{1000}{\mega\hertz} region, time domain Brillouin scattering \cite{klieber_optical_2012} in the \SIrange[scientific-notation = engineering]{1}{10}{\giga\hertz} region and Picosecond Ultrasonics, exciting longitudinal acoustic waves \cite{morath_phonon_1996} or shear acoustic waves \cite{pezeril_optical_2009} up to several \SI{100}{\giga\hertz}.
These high-frequency techniques are limited to rather high moduli and low loss as the sample must be able to support a sound wave at the given frequency.

The Piezo-electric Shear Gauge (PSG) \cite{christensen_rheometer_1995} is a rheometric method originally developed with a focus on measuring shear mechanical properties of supercooled liquids.
It is an electromechanical transducer that utilizes the coupling of strain and electric field in a piezo-electric material using piezo-electric ceramic (PZ) discs.
A sample attached to the PZ disc will experience a tangential force on its surface as the PZ disc deforms when a field is applied through electrodes on its surface.
If the sample is stiff, it will partially clamp the motion of the PZ disc, thus lowering its capacitance. The shear modulus of the sample can be calculated based on the capacitance of the PZ discs.
This makes the PSG ideal for relatively hard samples (moduli from \si{\mega\pascal} to tens of \si{\giga\pascal}) exactly because the deformation of the measuring device (the PZ discs) is the essence of the method.
This, however, also makes it less suitable for soft materials and low-viscosity liquids.

The PSG is a quasi-static method and has a uniquely broad frequency range (\SI{1}{\milli\hertz}-\SI{10}{\kilo\hertz}) due to the high resonance frequency of the device ($\sim \SI{100}{\kilo\hertz}$).
The method has been in use for more than two decades and has been pivotal in several 
scientific achievements: the development of a theoretical model for the non-Arrhenius temperature dependence of viscosity of supercooled liquids \cite{dyre_local_1996} and subsequent tests of that model \cite{eliasen_high-frequency_2021}, testing a prediction of the isomorph theory connecting the empirically found density scaling exponent to linear response measurements
\cite{gundermann_predicting_2011}, the first evidence of short-chain polymer-like rheological signatures in  monoalcohols \cite{gainaru_shear-modulus_2014, hecksher2014_supra} and a poly-alcohol \cite{jensen_slow_2018}, compiling the first true broadband mechanical spectrum in conjunction with high frequency propagation methods spanning  14 decades in frequency \cite{hecksher_toward_2017}, testing rheological models for the shear relaxation in supercooled liquids \cite{hecksher2017_squalane,hecksher2022_alphamodel}, showing that time scales from different response functions (including rheological) have the same temperature dependence \cite{jakobsen_communication_2012, roed_time-scale_2021} and showing that the local and global shear moduli are identical \cite{weigl_identity_2021}.

\begin{figure*}
	\includegraphics{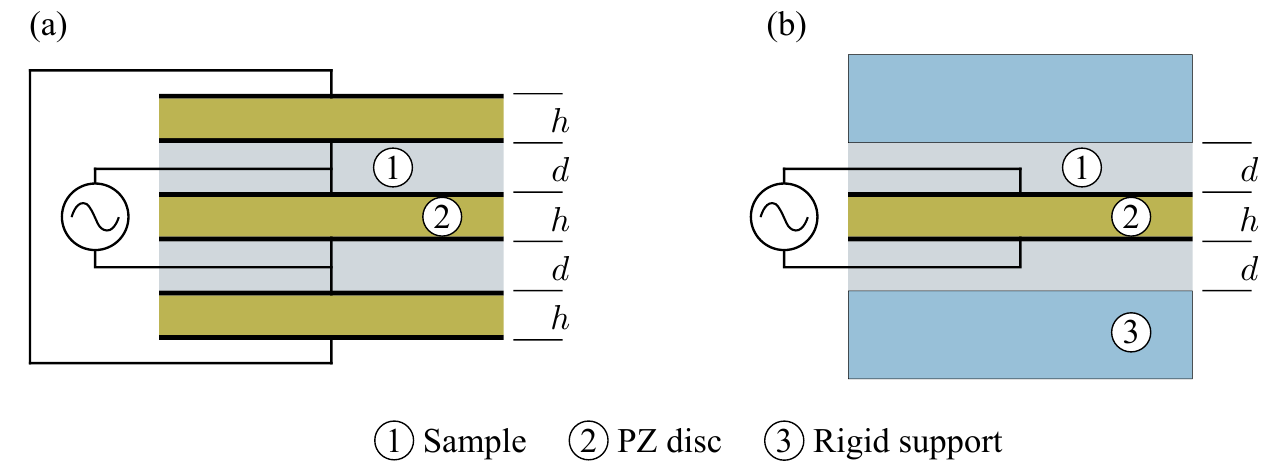}
	\caption{
		Schematic of the piezoelectric shear modulus gauge (PSG).
		a) 3-PZ version.
		Two layers of sample of thickness $d$, are sandwiched between three PZ discs of thickness $h$.
		Silvering on the surfaces of the PZ discs forms the electrodes of capacitors.
		The three capacitors are connected to an AC voltage source.
		The two outer PZ discs are connected in series, and the middle PZ disc is in parallel with the others.
		b) 1-PZ version.
		A layer of sample on either side of the PZ disc is sandwiched between the PZ and a rigid disc.
	    In the 3-PZ PSG, the electrodes on the sides facing each other of two neighbouring PZ discs are at the same potential and thus there is no electrical field across the sample. In the 1-PZ PSG, the field outside the capacitor formed by the silvered surface electrodes of the PZ disc is also zero everywhere except for a small field at the edge, which will be negligible due to the high dielectric permittivity of the PZ ceramic material. Thus, in neither case do the dielectric properties of the sample contribute to the signal.
	}
	\label{fig:PSG_Schematic}
\end{figure*}

The original PSG of \citet{christensen_rheometer_1995} consists of three PZ discs between which two layers of a liquid sample are suspended.
We will refer to this as the ``3-PZ PSG''.
In this work, we present a new version of the PSG where the two outer PZ discs have been replaced by rigid discs (made from sapphire or steel), such that the PSG consists of a single PZ disc mounted between two rigid supports.
This assembly will be referred to as the  ``1-PZ PSG''. Figure \ref{fig:PSG_Schematic} shows a schematic illustration of the 3-PZ and the 1-PZ PSG.

The 1-PZ PSG allows for use on sample types that the 3-PZ PSG design is not well-suited for.
Liquid samples are loaded into the 3-PZ PSG from the side assisted by capillary forces that pull the liquid into the sample space and ensure a homogeneously distributed liquid layer.
To mount solid samples, e.g. polymers and rubbers, a design is needed that can easily be disassembled so that discs of the sample material with matching radii can be sandwiched between the PZ discs/rigid support.
Solid samples need to be glued to the discs to ensure the no-slip condition that is essential for the method to work.
Because the PZ discs are quite brittle, they break easily when solid samples are removed after a measurement, and thus a 1-PZ PSG design reduces the potential waste of discs. 
Another advantage of using one instead of three PZ discs, is to avoid the tedious and time consuming task of matching three discs carefully among a batch of commercial ceramic discs. The matching is necessary, because the method ideally requires the three discs to be identical (same size, weight, capacitance, etc.).
In addition, in the 3-PZ assembly it is necessary to drill holes in the two outer discs, which again entails the risk of breaking the discs.

Another use case for the 1-PZ PSG is given by measurements on conductive samples, e.g. ionic liquids as was done by \citet{eliasen_high-frequency_2021}, where the risk of excess liquid at the edge of the PZ discs (which could short the electrodes) is greatly reduced in the 1-PZ design compared to the 3-PZ PSG.

Furthermore, we report on recent developments in the data extraction procedure that yield more precise results and extend the frequencies that can be resolved with this technique.

In this paper, we present the needed background and construction details for the 1-PZ PSG and show the application for glassy materials.
The paper starts with a brief review of the mathematical modelling of the PSG from \citet{christensen_rheometer_1995} in Sec.~\ref{sec:the_piezoelectric_shear_modulus_gauge} to give the background for the new data extraction algorithm described in Sec.~\ref{sec:inversion_algorithm}.
In Sec.~\ref{sec:measurements}, we validate the 1-PZ PSG against 3-PZ PSG measurements with a liquid sample in a fixed assembly with sapphire supports.
Interestingly, measurements with this cell initially gave slightly different results when compared to 3-PZ PSG results.
The slight deviations in the measured time scales are fully explained by taking into account the flow of a highly viscous liquid in geometrical confinement imposed by the PSG plates, and the problem was solved by introducing soft spacers in the 1-PZ PSG, allowing for thermal contraction of the liquid when cooled to temperatures near the glass transition.

\section{The Piezoelectric Shear Modulus Gauge}
\label{sec:the_piezoelectric_shear_modulus_gauge}
\begin{figure}
	\includegraphics{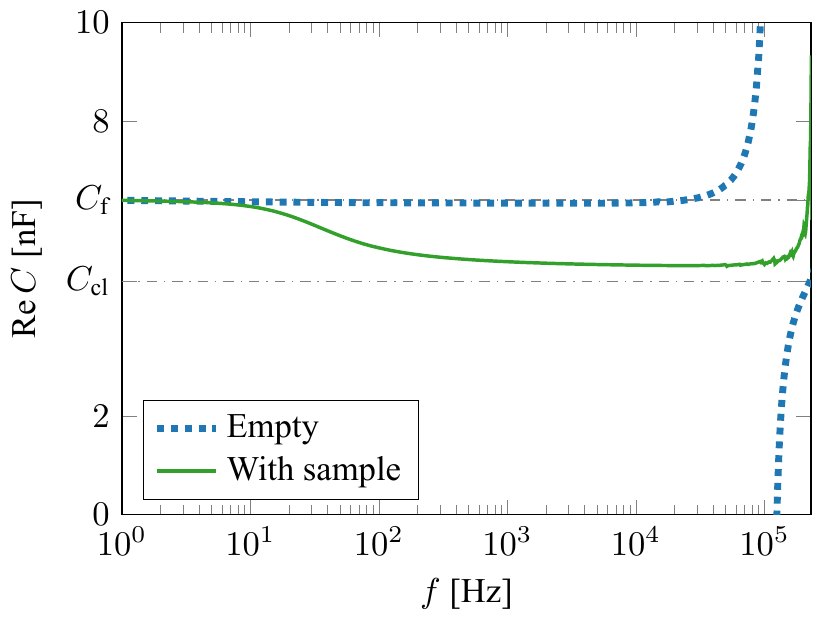}
	\caption{
		Real part of the capacitance of a 1-PZ PSG transducer as a function of frequency.
		Dotted lines are data from an empty PSG, and full lines are data with a liquid sample.	The PSG (without sample) has one capacitance, $C_\text{f}$, in the mechanically free state and another, lower, capacitance $C_\text{cl}$ in the mechanically clamped state. The dashed lines represent the calculated values.
		If a liquid sample is mounted, the PZ still moves freely at low frequencies, where the sample modulus approaches \SI{0}{\pascal}.
		At frequencies between this limit and the resonance, the sample modulus becomes significant.
		The sample then partially clamps the PZ discs, and the capacitance drops to a level between $C_\text{f}$ and $C_\text{cl}$.
		It is this partial clamping -- and the associated change in capacitance -- that enables us to measure the shear modulus using the PSG.}
	\label{fig:PSG_RawData}
\end{figure}

The PSG works by measuring the capacitance of the PZ discs as a function of frequency.
When an AC voltage is applied across the discs, they expand and contract in an oscillatory motion, thus shearing the sample sandwiched between the discs.
In the low-modulus limit (either at high temperatures or at low frequencies), the measured capacitance of the discs is the same as that of the freely moving discs, because the liquid is able to follow the imposed deformation without resistance.
If the sample is viscous, it resists the deformation and will partially clamp the discs, leading to a lower measured capacitance.
With a viscoelastic sample, the sample might flow at low frequencies and behave as an elastic solid at high frequencies, which shows up in the PSG as a transition between the free and partially clamped capacitance. Figure~\ref{fig:PSG_RawData} illustrates how the capacitance of the liquid filled PSG matches that of the empty device at low frequencies and decreases with frequency to the partially clamped level.
This difference in capacitance between the freely moving PZ disc and the partially clamped PZ disc is directly related to the shear modulus of the sample.
Through analysis of the equations of motion for the discs (briefly sketched in the following, for more details consult \citet{christensen_rheometer_1995}), the frequency dependence of the sample shear modulus can then be found by inversion of a function that maps sample modulus to PZ capacitance.

\subsection{Mapping of the PSG to ``one-one'' configuration}

In this section we show how both the 3-PZ PSG and the 1-PZ PSG can be mapped to a mathematically equivalent configuration that consists of one liquid layer between a PZ disc and an infinitely rigid support, which we will refer to as the ``one-one configuration''.
The mathematical model derived by \citeauthor{christensen_rheometer_1995}\cite{christensen_determination_1994} assumes this mapping, but the argument given here (and in App.~\ref{sec:mapping}) is more detailed. 
We establish that the model used for the 3-PZ PSG also applies for the 1-PZ PSG, only with a different effective sample layer thickness.

\begin{figure*}
	\includegraphics{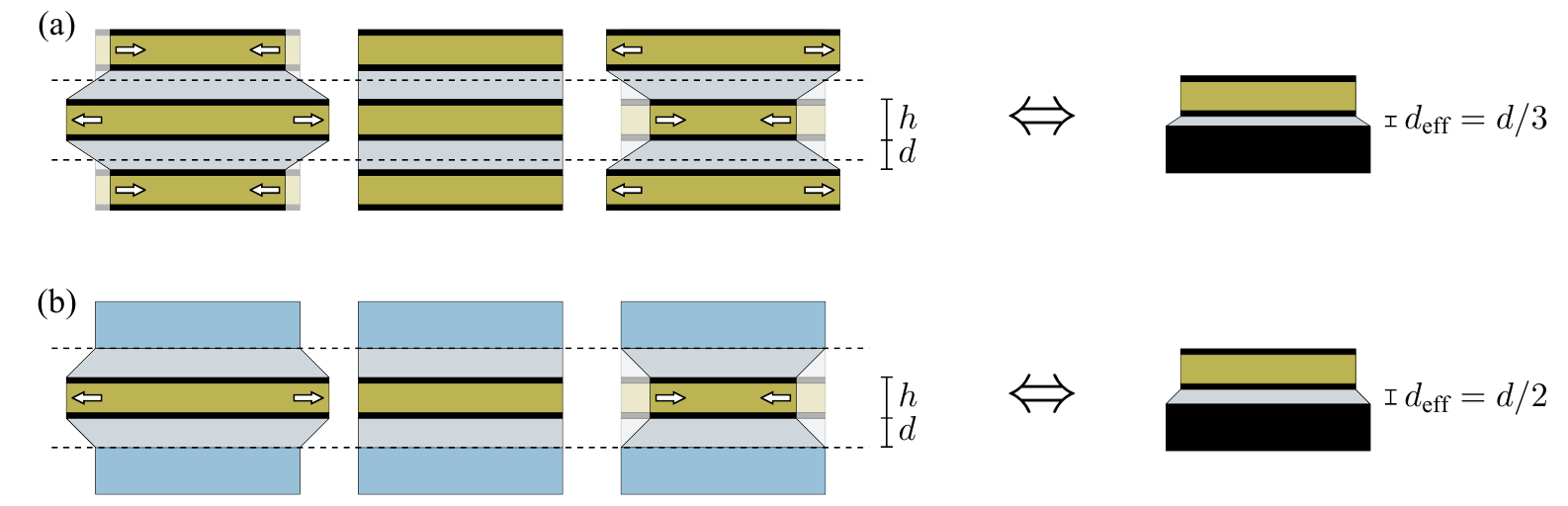} 
	\caption{
		Schematic illustration of the shearing motion of the two PSG implementations and their mapping to a one-one configuration.
		Arrows illustrate the direction of expansion/contraction of the discs.
		Note that these sketches highly exaggerate the disc displacement for illustrative purposes.
		The actual strain is on the order of \numrange{1e-5}{1e-6}.
		The ratio of radius and sample/disc thickness is also not to scale: in the measuring cells used in this work radius of the PZ disc is \SI{10}{\milli\metre} while $h \approx \SI{0.5}{\milli\metre}$ and $d \approx \SI{0.5}{\milli\metre}$ (see photo in Fig.~\ref{fig:PSG_Sapphire}).
		a) The 3-PZ PSG.
		This device, with a central PZ disc that moves in opposition to, and twice as far as, the outer PZ discs, has a sample thickness of $d$ and neutral planes (shown as dashed lines) that lie at one third of the distance from the outer discs to the central one.
		The 3-PZ PSG is therefore equivalent to a device with a single PZ, a sample thickness of $d/3$, and an infinitely rigid support below the sample.
		This equivalent device is shown on the right.
		b) The 1-PZ PSG.
		This has the neutral plane lying at the surfaces of the rigid discs.
		It therefore maps to a device with a sample thickness of $d/2$.
	}
	\label{fig:PSG_Schematic_Moving}
\end{figure*}

In the 3-PZ PSG, the middle disc moves opposite the two outer discs and always moves twice as much; both in the freely moving case and when there is a mechanical load of the sample.
See Fig.~\ref{fig:PSG_Schematic_Moving}(a) for an illustration.
This can be understood by considering an axial liquid filament reaching from the top plate to the middle plate.
If the filaments are non-interacting, then -- due to Newton's third law -- the stress one filament exerts on the top plate is equal and opposite to the stress it exerts on the middle plate.
This argument assumes the liquid inertia is negligible, which is reasonable in the quasi-static limit where the displacement accelerations 
are small.
As the frequencies approach the resonance, this assumption becomes worse. However, liquid inertia is accounted for by including a correction to the apparent modulus obtained by assuming negligible inertia (more details given below in Sec.~\ref{sec:the_piezoelectric_shear_modulus_gauge}\ref{sec:mapping} and App.~\ref{sec:Appinerti}).

Since the bottom plate exerts an equal stress on the middle plate, the total stress on the middle plate is twice the magnitude of, and opposite in direction to, each of the stresses on the outer plates
(see also Fig.~\ref{fig:PSG_ForceBalance}, App.~\ref{sec:mapping}).
By the electric wiring (see Fig.~\ref{fig:PSG_Schematic}), the voltage across the middle plate is also twice those across the outer plates, and thus the displacement of the middle plate is twice that of the outer plates, but in the opposite direction.
This means that there is a neutral plane $2/3$ of the distance from the middle plate both above and below the middle plate (see Fig.~\ref{fig:PSG_Schematic_Moving}).
This assumes that the sample displacement profile is linear, i.e.~the considered axial sample filaments are straight and not curved.
The assumption is reasonable, because in our case the strain is small ($10^{-5}$~to~$10^{-6}$) and as is the ratio of sample thickness $d$ and radius $R$ ($d/R = 0.05$).

Thus the outer plates ``see'' one rigid support at a distance of $(1/3) d$, illustrated by the dashed line in Fig.~\ref{fig:PSG_Schematic_Moving}(a).
The middle plate ``sees'' two rigid supports at a distance of $(2/3) d$, but this is equal to seeing only one rigid surface $(1/3) d$ away.
The three discs thus have exactly the same change in capacitance due to the sample shear modulus.
(A detailed mathematical derivation of this argument is given in App.~\ref{sec:mapping}.)
This means that the mathematical problem of relating the electrical capacitance of the 3-PZ PSG to the shear modulus of the sample can be mapped to the one-one configuration, consisting of a sample layer placed between a single PZ disc (with the full capacitance of the PSG) and an infinitely rigid support, the effective sample thickness being $1/3$ of the actual sample thickness. 

The 1-PZ PSG has actual physical rigid supports instead of a mathematically neutral plane, but the argument still holds if the support can be considered ``infinitely'' rigid, i.e.~that the stiffness of the support is much larger than the stiffness of the sample.
The mathematical models of the 3-PZ and the 1-PZ PSG are therefore the same, only with different effective sample thickness of $d/3$, respectively $d/2$, see Fig.~\ref{fig:PSG_Schematic_Moving}.

For this mapping to hold, a proper centering and parallelism of plate(s), sample, and supports in the implementation in the measuring cells is crucial. Details on how this is done practically in the two cells presented here are given in Sec.~\ref{sec:measurements}.

\subsection{Mathematical model of one-one configuration}
Having established the mapping to the one-one configuration, we now proceed to give the essential steps of the derivation of the mathematical model of that configuration.
Equations \eqref{eq:kp}-\eqref{eq:Fshear} are taken directly from \citeauthor{christensen_rheometer_1995}\cite{christensen_rheometer_1995}.
They are included here to give the complete background for the new inversion algorithm that determines the shear modulus.

The PSG (without sample) has one capacitance, $C_\text{f}$, in the mechanically free state and another, lower, capacitance $C_\text{cl}$ in the mechanically clamped state.
The free capacitance is equal to the low-frequency limiting capacitance, $C_0 $, i.e.~$C_0 = C_\text{f}$.
These levels, $C_\text{f}$ and $C_\text{cl}$, are illustrated in Fig.~\ref{fig:PSG_RawData}, that shows the measured capacitance as a function of frequency for an empty (blue dashed line) as well as a filled PSG (green full line).
The clamped and the free capacitances are related by the planar coupling constant, $k_\text{p}$,
\begin{equation}
	\frac{C_\text{cl}}{C_\text{f}} = 1 - k_\text{p}^2. \label{eq:kp}
\end{equation}
The ceramic material used in our experiment is a lead zirconate titanate compound known as Pz26.
The coupling constant, $k_\text{p}$, for this material is nominally \num{0.56}\cite{meggitt_as_data_2019}, and thus one expects $C_\text{cl}/C_\text{f} = 0.69$.
The actual value varies as a function of temperature and thermal history, and thus it is calibrated for each measurement. Other characteristics of the Pz26 material include a density of \SI{7.70}{\gram\per\centi\metre\cubed} and a relative dielectric permittivity of 1300 at \SI{1}{\kilo\hertz}\cite{meggitt_as_data_2019}.

In the piezoelectric material the stress, $\sigma_{ij}$, and strain, $\epsilon_{ij}$, couple to each other as well as to the electric field, $E_i$, and the displacement field, $D_i$.
For an axially polarized and cylindrically symmetric PZ disc, these couplings can be reduced to\cite{christensen_rheometer_1995}
\begin{align}
	\begin{pmatrix}
		\sigma_{r r} \\ \sigma_{\phi \phi} \\ D_z
	\end{pmatrix}
	&=
	\begin{pmatrix}
		c_{11} & c_{12} & -e_{13} \\
		c_{12} & c_{11} & -e_{13} \\
		e_{13} & e_{13} & \varepsilon^S_{33}
	\end{pmatrix}
	\begin{pmatrix}
		\epsilon_{r r} \\ \epsilon_{\phi \phi} \\ E_z
	\end{pmatrix},
\end{align}
where $c_{11}$ and $c_{12}$ are elastic constants, $\varepsilon^S_{33}$ is a dielectric constant and $e_{13}$ is a piezoelectric constant.
$r$ denotes a radial component, $z$ a component in the axial direction, and $\phi$ an azimutal component.

The measured capacitance of the PZ disc, $C_\text{m}$, can be found as the ratio of charge, $Q$, to voltage, $U$, with the charge found from the surface integral of the displacement field, $D_z$, and the voltage found from the electric field and the thickness, $h$, of the PZ disc,
\begin{align}
	C_\text{m} &= \frac{Q}{U} = \frac{\int_0^{R} 2 \pi r D_z(r) \dif r}{h E_z}.
\end{align}
This leads to the capacitance only being a function of the radial displacement at the edge of the PZ disc, $u_r(r = R)$,
\begin{align}
	C_\text{m} &= \frac{2\pi e_{13} R}{h E_z} u_r(R) + C_{\text{cl}}\,.
\end{align}

The pivotal function describing the shear transducer response is thus the normalized capacitance,
\begin{equation}
	F = \frac{C_\text{m} - C_\text{cl}}{C_\text{f} - C_\text{cl}}\,,\label{eq:Fdef}
\end{equation}
which is directly proportional to the displacement, $u_r(R)$.

The displacement may be found by solving the equation of motion of the PZ disc, which in cylindrical coordinates becomes
\begin{align}\label{eq:diffeq}
	c_{11} \left( r^2 u_r^{\prime\prime} + u_r^\prime r - u_r \right) - \frac{\sigma_l}{h} r^2 &= - \omega^2 r^2 \rho u_r,
\end{align}
where a prime denotes a derivative in $r$, $\omega$ is the frequency of the electrical field and $\rho$ is the density of the ceramic.
$\sigma_l$ denotes the tangential stress exerted by the sample on the PZ disc and is proportional to the shear modulus, $G(\omega)$, of the sample, $\sigma_l = G(\omega){u_r(r)}/{d_\text{eff}}$.
At high frequencies and low moduli one has to take liquid inertia into account \cite{schrag_deviation_1977}.
This can be done by substituting $G(\omega)$ by the apparent modulus $G_\text{app}(\omega) = G(\omega) \frac{x}{\tan(x)}$, where $x = \sqrt{\rho_l/G(\omega)}\omega\,d$ (see App.~\ref{sec:Appinerti} for more details).
Rewriting the equation of motion Eq.~\eqref{eq:diffeq} in terms of the dimensionless variable $x = r/R$, we obtain
\begin{equation}\label{eq:diffeq2}
	x^2 e^{\prime\prime} + x e^\prime +\left( \left[ \frac{\omega^2\rho R^2}{c_{11}} - \frac{G(\omega)R^2}{c_{11} d_\text{eff} h} \right] x^2 -1 \right) e = 0,
\end{equation}
where $e$ is now the normalised displacement (see App.~\ref{sec:including_effects_of_partial_liquid_filling_and_the_hub_in_the_psg}).

Defining the characteristic frequency, $\omega_c$, and modulus, $G_\text{c}$, of the PSG as
\begin{equation}\label{eq:omegac}
	\omega_c^2 =\frac{c_{11}}{\rho R^2} ,\quad G_\text{c} = \frac{c_{11} d_{\text{eff}} h}{R^{2}},
\end{equation}
it becomes clear that $F$ depends on the frequency and the shear modulus $G(\omega)$ via the wave vector, $k$,
\begin{equation}\label{eq:ksquare}
	k^2 = \left(\frac{\omega}{\omega_\text{c}}\right)^2 - \frac{G(\omega)}{G_\text{c}}.
\end{equation}

Solving the equation of motion Eq.~\eqref{eq:diffeq2} with the appropriate set of boundary conditions, we obtain for the normalized capacitance (for more details, see \citet{christensen_rheometer_1995})
\begin{equation}
	F(k) = (1 + \nu) \frac{J_1(k)}{k J_0(k) + (\nu - 1) J_1(k)}, \label{eq:Fshear}
\end{equation}
where $J_0$ and $J_1$ are Bessel functions and $\nu \approx 0.31$is the Poisson ratio of the Pz26 material\footnote{in \citeauthor{christensen_rheometer_1995} the symbol $p$ was used in stead of $\nu$. We change it here to be consistent throughout the current paper.}.

The above model of the PSG has not involved any dissipation, and thus the imaginary part of the capacitance is zero.
In practice, a peak is seen in the imaginary part at resonances due to a small dissipation.
This is included simply by adding a quality factor $Q_\text{pz}$ into the expressions for $k^2$, i.e.
\begin{align}\label{eq:kdamped}
	k^2 &= \left(\frac{\omega}{\omega_\text{c}} \right)^2 - i \left( \frac{\omega}{\omega_\text{c}} \right) \frac{1}{Q_\text{pz}} - \frac{G(\omega)}{G_\text{c}}.
\end{align}
Other adjustments to the model include
1) a correction for thermal contraction of the sample (see App.~\ref{sec:including_effects_of_partial_liquid_filling_and_the_hub_in_the_psg}) resulting in $F$ also being a function of the actual sample radius, $x_l$, $F = F(k,x_l)$,
2) a correction for the hub in the centre of the PSG (see  App.~\ref{sec:details_on_the_inelastic_hub_calculations}),
and 3) an assumption to handle the weak dispersion of the ceramic material, i.e.~that $C_\text{f}$ and $C_\text{cl}$ are not constants, but have a weak frequency dependence.
To account for this frequency dependence, we assume that their ratio, $C_\text{cl}/C_\text{f}$, -- and thus the coupling constant (Eq.~\eqref{eq:kp}) -- is frequency independent.
In that case, the effect of dispersion can be scaled out by a reference measurement of the empty transducer, $C_r$.
Taking the ratio of the reference and sample spectra then scales out the dispersion.
The ratio becomes
\begin{equation}
	\frac{C_\text{m}}{C_\text{r}} = \frac{F(S,V,x_l) \frac{k_\text{p}^2}{1 - k_\text{p}^2} + 1}{F(S,0,1) \frac{k_\text{p}^2}{1 - k_\text{p}^2} + 1},
\end{equation}
where we have introduced the notation
\begin{equation}
	S = \left( \frac{\omega}{\omega_\text{c}} \right)^2 - i \left( \frac{\omega}{\omega_\text{c}} \right) \frac{1}{Q_{\text{pz}}}, \quad
	V = \frac{G(\omega)}{G_\text{c}},
\end{equation}
replacing the previous expression for $k^2$, Eq.~\eqref{eq:kdamped}, by $k^2 = S - V$.

Thus the normalized capacitance for the sample measurement may be written as
\begin{equation}\label{eq:Ffromref}
	F(S,V,x_l) = \frac{C_\text{m}}{C_\text{r}} \left( F(S,0,1) + \frac{1 - k_\text{p}^2}{k_\text{p}^2} \right) - \frac{1 - k_\text{p}^2}{k_\text{p}^2}\,,
\end{equation}
where the right-hand side is calculated from the experimentally determined $C_\text{m}$ and $C_\text{r}$, while $V$ is found by inversion of $F(S,V,x_l)$.

\section{New Inversion Algorithm}
\label{sec:inversion_algorithm}
The model gives the ingredients to determine the frequency-dependent shear modulus of the sample from a measurement of the capacitance of the PZ discs. The characteristic properties of the PSG ($C_\text{cl}$, $k_\text{p}$ and $\omega_\text{c}$) can in principle be determined by the reference measurement of the capacitance, $C_r$.

However, it is not possible to isolate $G(\omega)$ in the expression for $F(k)$ Eq.~\eqref{eq:Fshear} (or more generally Eq.~\eqref{eq:Fshear_rh}, see App.~\ref{sec:details_on_the_inelastic_hub_calculations}) in a simple way.
The strategy in \citet{christensen_rheometer_1995} involved approximating the function
\begin{equation}
	\Phi(S,V,x_l) = \frac{F(S,V,x_l)}{F(S,0,1)}
\end{equation}
with a rational function with a numerator of degree one and a denominator of degree two in $G(\omega)/G_\text{c}$.
The inversion to isolate $G(\omega)$ was based on this expression.

$F(S,V,x_l)$ may also be directly inverted by fitting $G(\omega)$ for each frequency in Eq.~\eqref{eq:Ffromref}.
This method is accurate, but time consuming when implemented with a non-linear, least-squares brute-force algorithm, since a minimisation is executed for each measured frequency ($\sim 150$) in the spectrum.

In the following we present an implementation of the inversion using the Newton-Raphson algorithm \cite{Ryabenkii2006}, which yields results identical to those obtained with the brute-force algorithm, but is much faster.

Specifically, we use the Newton-Raphson algorithm to determine the zero-points of the difference between the measured normalized capacitance, $F_m$, and the theoretical expression for $F$ in Eq.~\eqref{eq:Fshear} (or more generally Eq.~\eqref{eq:Fshear_rh}) with the sample shear modulus, $G(\omega)$, as the variable.
The Newton-Raphson inversion algorithm should thus converge to the solution, $G$, of the equation
\begin{equation}
	F_\text{m} = F(G),
\end{equation}
where $G$ is the limit of the iterations of trial functions $G_\text{t}^{(n)}$ and
\begin{equation}
	G_\text{t}^{(n + 1)} = G_\text{t}^{(n)} \left( 1 + \delta \frac{F_\text{m} - F\left(G_\text{t}^{(n)}\right)}{F\left((1 + \delta) G_\text{t}^{(n)}\right) - F\left(G_\text{t}^{(n)}\right)} \right).
\end{equation}
$\delta$ is a small number (typically \num{0.01}) that sets the step size of the numerical differentiation of $F$.

\begin{figure}
	\includegraphics{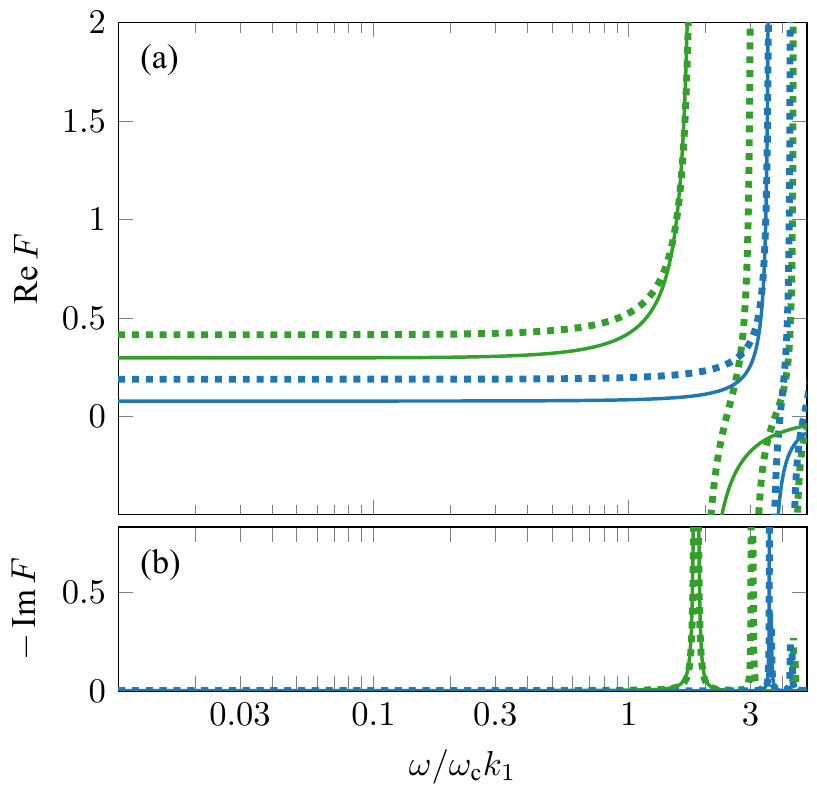}
	\caption{
		The normalized capacitance, $F$ (dotted lines), of the PSG transducer as a function of frequency along with the trial function, $F_1$ (solid lines), for two values of
		$G(\omega)$: $G(\omega) = (10 + 0.1i) G_\text{c}$ (green) and $G(\omega) = (50 + 0.1i) G_\text{c}$ (blue).
		The frequencies are scaled relative to the resonance frequency of the empty transducer, $\omega_\text{c} k_1$, where $\omega_\text{c}$ is the characteristic frequency of the PSG and $k_1 = 2.054$.
	}
	\label{fig:PSG_Functions}
\end{figure}

\begin{figure*}
	\centering
	\includegraphics{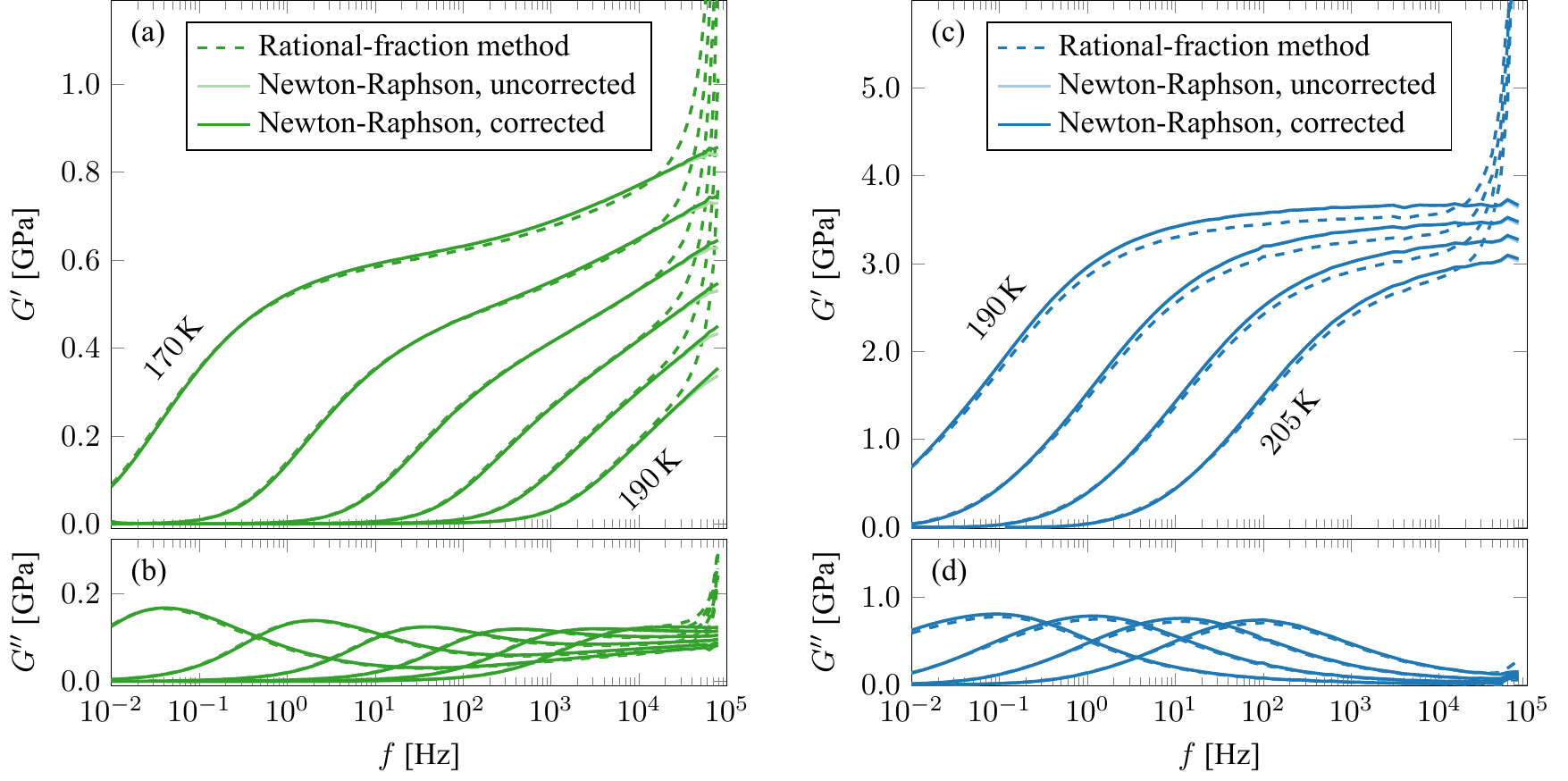}
	\caption{
		Comparison of inversion methods used on data from squalane (left) and glycerol (right).
		These data are obtained with the 3-PZ PSG.
		To compare the two different inversion algorithms, we have used raw data (i.e.~measured capacitance of the PSG with and without liquid) and inverted them using the two algorithms.
		Shear moduli extracted from these data have previously been published by \citet{roed_time-scale_2021} (squalane) and \citet{jensen_slow_2018} (glycerol).
		Dashed lines are data inverted using the algorithm described in \citet{christensen_rheometer_1995}.
		Lighter colored solid lines are inverted using the Newton-Raphson algorithm, the darker solid lines have subsequently been corrected for sample inertia.
		The difference between corrected and uncorrected data is only visible at frequencies above \SI{30}{\kilo\hertz} and only in the storage modulus for squalane.
		The glycerol moduli are so high that the correction is almost not visible.
		Note how data inverted by the rational fraction algorithm diverges as the frequency approaches the resonance frequency of the PSG (at \SI{100}{\kilo\hertz}), while data inverted by the new algorithm levels off to approach $G_\infty$, thus extending the frequency range up to \SIrange[range-phrase=--]{50}{70}{\kilo\hertz}.
	    }
	\label{fig:Inversion_Comparison}
\end{figure*}

For the method to work, we need a good starting trial function, $G^{(1)}_\text{t}(\omega)$.
To obtain this, the normalized capacitance of the shear transducer, $F(k)$, will be approximated by a rational fraction that matches the measured $F_\text{m}$ up to a frequency a little above the first resonance frequency.
$F$ has its first resonance, $\omega_1 = k_1\omega_\text{c}$, for $k_1 = 2.054$ when $p = 0.31$ (see \citet{christensen_rheometer_1995}).
Thus, by measuring $\omega_1$, we can find $\omega_\text{c}$ and $G_\text{c}$ of Eq.~\eqref{eq:omegac}.
The function
\begin{equation}
	F_1(k) = \frac{1}{1 - \left( \frac{k}{k_1} \right)^2} \label{eq:F1}
\end{equation}
matches $F(k)$ both for $k\rightarrow 0$ and at the resonance frequency.
Inverting $F_1$ using the non-dissipative definition of $k$ (Eq.~\eqref{eq:ksquare}) gives
\begin{equation}
	G_\text{t}^{(1)} = G_\text{c} \left[ \left( \frac{\omega}{\omega_\text{c}} \right)^2 + k_1^2 \frac{1 - F_1}{F_1} \right], \label{eq:G1}
\end{equation}
which may then be used as an initial trial function for the Newton-Raphson inversion algorithm with $F_\text{m}$ inserted for $F_1$.

Figure~\ref{fig:PSG_Functions} shows model examples of $F$ and $F_1$ for cases where the sample modulus is set to values of $G = (10 + 0.1i) G_\text{c} \sim \SI{0,7}{\giga\pascal}$ and $G = (50 + 0.1i) G_\text{c} \sim \SI{3,5}{\giga\pascal}$, corresponding to ``soft'' and ``hard'' liquids, respectively.
The shear modulus of a hard liquid like glycerol is $\sim\SI{4}{\giga\pascal}$\cite{schroter_viscosity_2000,scarponi_brillouin_2004,klieber_mechanical_2013}, which is a little more than $50$ times the characteristic modulus, $G_\text{c} = \SI{0.07}{\giga\pascal}$, of the PSG.

Figure~\ref{fig:Inversion_Comparison} presents a comparison between the previous inversion algorithm\cite{christensen_rheometer_1995} for determining $G(\omega)$ of the sample and the Newton-Raphson implementation.
Figure~\ref{fig:Inversion_Comparison}(a+b) shows data on squalane, a ``soft'' liquid, which in a addition to the main (alpha) relaxation feature a secondary (beta) contribution ($G_\infty < \SI{1}{\giga\pascal}$ \cite{hecksher_toward_2017}).
Figure~\ref{fig:Inversion_Comparison}(c+d) shows the same comparison for glycerol, which has a high plateau modulus ($G_\infty \sim \SI{4}{\giga\pascal}$ \cite{jensen_slow_2018}).
The results from the previous procedure and the Newton-Raphson algorithm yield comparable results in the low-frequency region.
At higher frequencies, though, the Newton-Raphson method clearly gives better results.
The rational-fraction algorithm introduces an artefact where both real and imaginary parts of the modulus rise sharply as the frequency approaches the resonance, which is located around \SI{100}{\kilo\hertz}.
Moduli computed with the Newton-Raphson method approach a plateau, $G_\infty$ at high frequencies.
As expected, the inversion cannot go above the first resonance frequency of the empty transducer.

\section{Results and discussion}
\label{sec:measurements}
In this section we present measurements carried out with the 1-PZ PSG.
All measurements are carried out using cryostats and measurement equipment constructed at Roskilde University \cite{igarashi_cryostat_2008, igarashi_impedance-measurement_2008}.

\begin{figure}
	\includegraphics{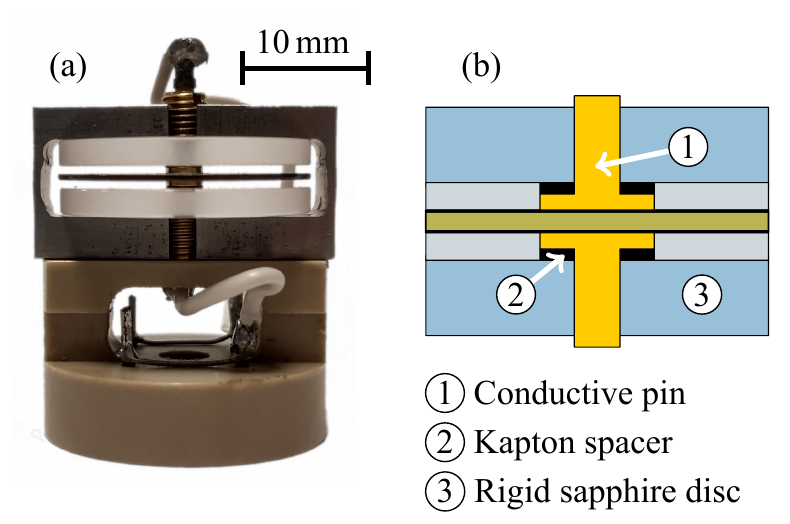}
	\caption{
		a) Photo of 1-PZ PSG for liquid samples.
		The rigid supports are made from sapphire glass and are held in place by an aluminium casing.
		Conductive pins leading through the sapphire glass hold the PZ disc in place and provide electrical connection to the electrodes.
		These electrode pins also define the thickness of the sample space and their presence means that this PSG has a rigid hub in the centre (see section~\ref{sec:measurements}.\ref{sec:platebending}).
		Wires lead from the pins to connecting points for measurement equipment.
		b) Schematic drawing of the centre of the device.
		Kapton spacers were placed between the flat head of the electrode pin and the sapphire glass to allow the liquid sample to contract axially.
	}
	\label{fig:PSG_Sapphire}
\end{figure}

Figure \ref{fig:PSG_Sapphire} shows a photo and a schematic drawing of the 1-PZ PSG implementation designed for liquid measurements (for a different implementation intended for solid samples, see App.~\ref{sec:rubber}).
The rigid outer discs are made from sapphire glass.
When assembling the cell, the two sapphire discs are self-centered in the cell casing by the electrode pins that go through a small hole drilled in the centre of the thick sapphire discs.
The PZ disc in the middle is then clamped between the ends of the two electrode pins, which have flat heads ensuring electrical contact, but also acting as spacers.
These spacers ensure parallelism of the plates.
The horizontal alignment of the middle PZ disc is done by hand when assembling and later inspected in a microscope.

The sample is loaded into the cell from the side at room temperature where the liquid has a low viscosity and is drawn into the gaps by capillary forces.
It is subsequently cooled to temperatures close to the glass transition temperature ($\sim$ \SI{210}{\kelvin}) covering viscosities from \SI{1}{\kilo\pascal\second} to  \SI{100}{\giga\pascal\second}.

As argued above, the 1-PZ PSG should yield results identical to those of the 3-PZ PSG if the assumption holds that the rigid discs replacing the outer PZ discs can be considered "infinitely" rigid.
We test this assumption by comparing two consecutive measurements in the same experimental setup (same cryostat, same electronics):
One with a 3-PZ PSG and one with a 1-PZ PSG.
Results obtained with the 3-PZ PSG have previously been found to be in agreement with other mechanical measurements: \citet{gainaru_shear-modulus_2014} used master curves obtained by standard rheological measurements and spectra from a 3-PZ PSG to evidence a low-frequency rheological signal in mono-alcohols like that of a short chain polymer, and  \citet{hecksher_toward_2017} combined PSG spectra with ultra high-frequency propagation methods form true broadband mechanical spectra spanning 14 decades in frequency.

The sample chosen for these measurements is tetramethyl-tetraphenyl-trisiloxane (DC704), a diffusion-pump oil used as a model glass-former\cite{niss_perspective:_2018}.
DC704 is chemically stable, it does not absorb water, and it has been measured numerous times\cite{niss_dielectric_2005,jakobsen_dielectric_2005,hecksher_relaxation_2011,hecksher_mechanical_2013} showing that it obeys time-temperature superposition and that its mechanical and electrical properties do not change over time.
Thus DC704 is ideal for this test, because any discrepancy between data obtained by 3-PZ and by 1-PZ measurements must originate from the cell and not the sample.

\begin{figure*}
	\includegraphics{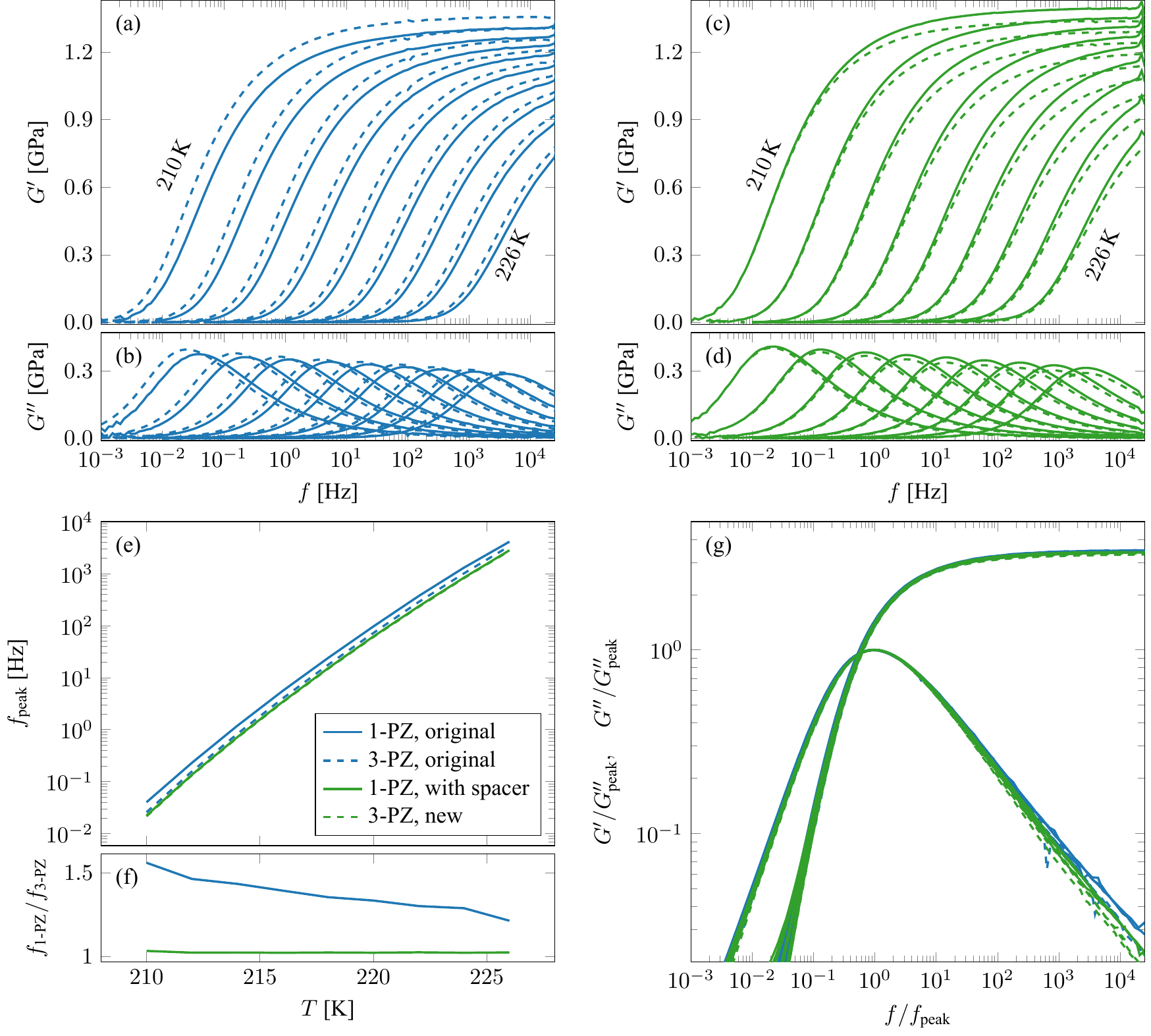}
	\caption{
		Shear modulus of DC704, comparing two sets of measurements made using a 3-PZ PSG and a 1-PZ PSG (a total of four measurements: both 3-PZ and 1-PZ PSG measurements were repeated, since the two sets were carried out in different cryostats with slightly different absolute temperature calibration).
		Measurements were made at temperatures from \SIrange{210}{226}{\kelvin} in steps of \SI{2}{\kelvin}.
		a) and b) show the storage modulus, \Gprime, and loss modulus, \Gdoubleprime, respectively, measured using a 3-PZ PSG and a 1-PZ PSG (without a soft spacer).
		c) and d) show \Gprime and \Gdoubleprime, measured using a 3-PZ PSG and a 1-PZ PSG with kapton spacers that allow the sample to contract axially.
		e) shows the loss peak position as a function of temperature for all four measurements.
		f) shows the ratio of loss peak position, $f_\text{peak}$, of the 3-PZ and the 1-PZ PSG measurements, respectively. Clearly, the time scales obtained with the 1-PZ PSG are different than those obtained with the 3-PZ PSG in the original measurements, and the ratio is increasing as the temperature is lowered. Identity of the time scales is restored, when a soft spacer is added in the 1-PZ PSG.
		g) shows all spectra scaled on the frequency axis with the loss peak frequency and modulus (both real and imaginary part) scaled by the value of $G''$ at the loss peak, $G''_{\text{peak}}$.
		Plotted this way, all four data sets collapse within the experimental uncertainty, showing that the sample obeys time-temperature superposition (TTS), i.e.~the spectral shape is unchanged between the different measurements.
		The deviations seen at the flanks of the loss peak are on the order of a few percent and reflect the uncertainty of the measurement.}
	\label{fig:DC704_Comparison}
\end{figure*}

The result of the initial measurements are shown in Fig.~\ref{fig:DC704_Comparison}(a+b).
The spectra from the 1-PZ (full lines) have a slightly different overall vertical scaling than the 3-PZ measurement (dashed lines).
This is a well-known uncertainty of the measurement of \SIrange{5}{10}{\percent}\cite{hecksher_relaxation_2011}.
However, a vertical scaling does not correct or explain the general shift of the spectra obtained with 1-PZ PSG to higher frequencies.

To explain this discrepancy, we look into the details of the thermal contraction of the sample in the two different cells upon cooling.
The liquid is filled into the cell at room temperature and thus it contracts when cooled towards its glass transition temperature.
For DC704, the isobaric thermal expansion coefficient is roughly $\alpha_P = \SI[exponent-to-prefix=false]{7e-4}{\per\kelvin}$, and the glass transition temperature is $T_g = \SI{210}{\kelvin}$.
This gives an estimated relative change in sample volume of $\sim \SI{5}{\percent}$ at the measurement temperatures.

One would guess that this contraction happens primarily in the radial direction, where the surface is free.
However, as the liquid cools, its viscosity increases enormously.
Consequently, the radial flow of the contracting liquid is severely hampered due to the confined geometry, where the sample thickness is much smaller than the radius, $d \ll R$.
This was discussed by \citet{niss_dynamic_2012}, where the radial flow time, i.e.~the characteristic time for flow in the radial direction, was estimated to be $\tau_\text{flow} \propto \tau_M (R/d)^2$, where $\tau_M=\eta/G_\infty$ is the Maxwell relaxation time.
This estimate is based on a situation, where the liquid is not able to contract/flow in the axial direction, which can be assumed to be the case for the 1-PZ PSG with thick sapphire supports.
Thus at some finite temperature the radial flow time exceeds the experimental time scale and flow ceases.
Upon continued cooling, the liquid volume will then no longer change and the thermodynamic boundary condition ends up being closer to isochoric than isobaric.

However, in the 3-PZ PSG the outer PZ discs are relatively thin and flexible, so the flow might progress slightly differently due to bending of the outer plates. 
In the following subsections we show model calculations of the bending of a flexible disc in contact with liquid that is radially clamped and how that leads to a flow time that scales as $(R/d)^6$, indeed confirming that a liquid sample in the 3-PZ PSG is able to bend the outer discs during thermal contraction and that this is the main mechanism for volume change.

\subsection{Plate bending after stopped radial flow}

Consider a sample  between two cylindrical discs of equal radii, $R$.
The lower disc is thick and rigid, whereas the upper disc, of thickness $h$, is considered thin ($h \ll R$) and flexible with flexural rigidity $D$. 
The two discs are connected by a central hub of radius $r_\text{h}$. The gap between the discs is filled with liquid to the edge at ambient temperature.

We start out by analysing what happens if we assume that radial flow has stopped to find the characteristic length scale of bending.

Plate bending, $\zeta(r)$, due to a normal force per area, $p(r)$, is governed by the fourth-order differential equation \cite{landau_theory_1986} 
\begin{equation}\label{eq:LL752}
	D \laplacian^2 \zeta - p = 0,
\end{equation}
where $D$ is the flexural rigidity given by the Young's modulus, $E$, the Poisson ratio, $\nu$, and the thickness, $h$, of the plate,
\begin{equation}
	D = \frac{E h^3}{12 (1 - \nu^2)}.
\end{equation}
$\laplacian$ is the Laplace operator which in the cylindrically symmetric situation we consider here is given by
\begin{equation}
	\laplacian = \frac{1}{r} \dod{}{r} r \dod{}{r}.
\end{equation}

The radius variable, $r$, ranges between the fixed hub, at $r_\text{h}$, and the rim, at $R$.
The boundary conditions at $r_\text{h}$ are no displacement and no bending,
\begin{eqnarray}
	\zeta = 0, \label{eq:bnd1} \\
	\dod{\zeta}{r} = 0,\label{eq:bnd2}
\end{eqnarray}
and at $R$ the boundary conditions are
\begin{eqnarray}
	\dod{}{r} \laplacian \zeta = 0,\label{eq:bnd3} \\
	\laplacian \zeta - \frac{1 - \nu}{r} \dod{\zeta}{r} = 0.
\end{eqnarray}

\begin{figure}
	\centering
	\includegraphics{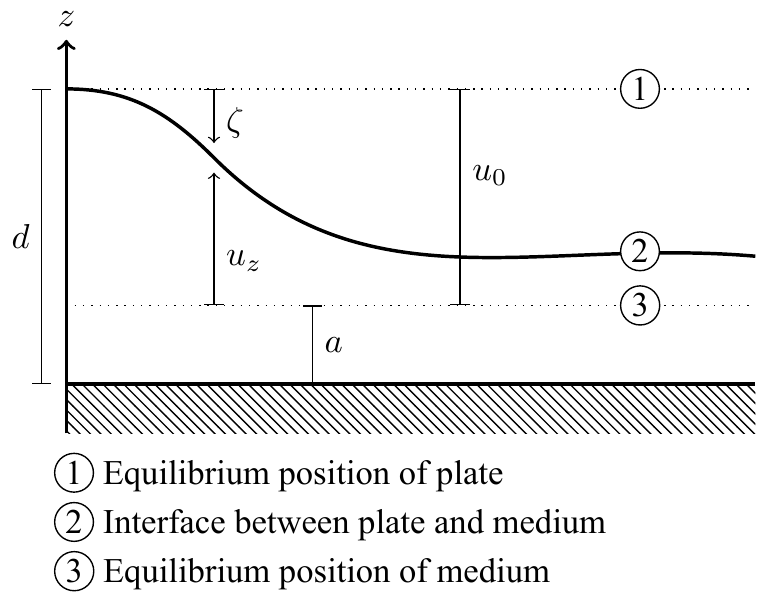}
	\caption{
		Plate bending by thermal contraction of a liquid in the axial direction.
		$\zeta$ is the deviation of plate position from equilibrium, $d$ is the distance between plate and rigid support and $u_0$ is the distance from plate equilibrium to elastic medium equilibrium.
	}
	\label{fig:Platebending_Schematic}
\end{figure}

The situation is depicted in Fig.~\ref{fig:Platebending_Schematic}.
The plate is in contact with an elastic medium with longitudinal modulus $M_T$.
The medium has an equilibrium thickness $a$, which before contraction of the liquid is equal to $d$.
$d$ is the distance between the rigid support and the flexible plate in the unbent state.
Upon cooling, the medium contracts by flow in the radial direction while $a$ stays equal to $d$.
However, at some temperature flow becomes hindered by the increasing viscosity and the medium starts to contract in the $z$ direction.
Then $a < d$ and the medium exerts an elastic force on the plate proportional to $u_z$, the displacement of the medium surface from the equilibrium position.
The force per area on the plate due to the elastic medium thus becomes
\begin{equation}\label{eq:key3}
	p = -\frac{u_z}{a} M_T,
\end{equation}
where $M_T$ is the longitudinal modulus.
Since $u_z = \zeta + u_0$ and $a \approx d$, we have
\begin{equation}\label{eq:platedif}
	\laplacian^2 \zeta + (\zeta + u_0) \frac{M_T}{Dd} = 0.
\end{equation}

\begin{figure}
	\centering
	\includegraphics{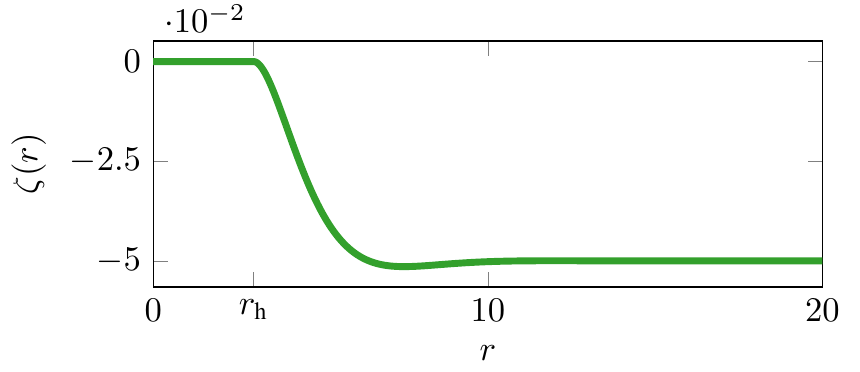}
	\caption{
		Solution to the differential Eq.~\eqref{eq:platedif3} governing bending of a thin disc, showing the radial profile of the disc, where $r_\text{h}$ is the radius of the hub (the head of the conductive pin in Fig.~\ref{fig:PSG_Sapphire}).
		Both axes are in units of the characteristic elastic length $l$ (see Eq.~\eqref{eq:charlength}).
		Since $l \sim d$, a bending of $\zeta = -0.05$ corresponds to the liquid contracting \SI{5}{\percent} in the axial direction. The radius of $20d$ thus corresponds to $\sim \SI{10}{\milli\metre}$ which is the radius of the PZ discs used in the measurements. 
		The profile is rather flat except close to the hub.
	}
	\label{fig:Platebending_Solution}
\end{figure}

Now we define a characteristic length
\begin{equation}\label{eq:charlength}
	l = \left( \frac{d D}{M_T} \right)^{1/4} = \left( \frac{d h^3 E}{12(1 - \nu^2)M_T} \right)^{1/4}.
\end{equation}
If $d \approx h$ and $E \approx M_T$ then $l \approx d$. We now use $l$ as the unit of length.
Then the differential Eq.~\eqref{eq:platedif} becomes
\begin{equation}\label{eq:platedif2}
	 \laplacian^2 \zeta + \zeta + u_0 = 0,
\end{equation}
which is an inhomogeneous differential equation.
If we however look at $u_z$ instead, we get the corresponding homogeneous equation which is more convenient to solve,
\begin{equation}\label{eq:platedif3}
	\laplacian^2 u_z + u_z = 0.
\end{equation}
Only the first boundary condition for $u_z$ differs from those of $\zeta$, becoming
\begin{equation}\label{eq:bnd4}
	u_z(r_\text{h}) = u_0.
\end{equation}
The solution to Eq.~\eqref{eq:platedif3} is a linear combination of the four Kelvin functions, Ber, Bei, Ker and Kei,
\begin{equation}\label{eq:bendingsolution}
	u_z(r) = A_1 \mathrm{Ber} (r) + A_2 \mathrm{Bei}(r) + A_3 \mathrm{Ker}(r) + A_4 \mathrm{Kei}(r),
\end{equation}
and the values of the four constants may be found by applying the boundary conditions (see App.~\ref{sec:platebending}).
Figure~\ref{fig:Platebending_Solution} shows an example of such a solution for the profile of the bending.

\subsection{Flow time}
We proceed to derive the characteristic flow time for the process where the radial pressure gradient exerted on the liquid by the bending plate makes the liquid flow.
Let $V(r)$ be the volume of the liquid inside radius $r$ and $\dot V(r)$ its time derivative.
In order to simplify the modelling of the flow we assume that different parts of the liquid flow \emph{either} in the radial direction \emph{or} in the axial direction.
The gain in volume between $r$ and $r + \dif r$ due to flow in the radial direction in time $\dif t$ is $\dot V(r + \dif r) \dif t - \dot V(r) \dif t$.
This must equal the loss in volume due to flow in the axial direction which is $\mmnote{-}2 \pi r \dif r \dot \zeta \dif t$.
The volume continuity equation thus becomes
\begin{equation} \label{eq:zetadot}
	\dot \zeta(z) = -\frac{1}{2 \pi r} \dod{\dot V}{r}.
\end{equation}
The flow in the $r$ direction is considered to be planar Poiseuille flow, implying that the one-dimensional volume flow per length orthogonal to the flow direction is $-\od{p}{r} d^3/12 \eta$.
If for the radial flow we take the length as $2\pi r$, we get
\begin{equation} \label{eq:planpoiseuille}
	\dot V = - \dod{p}{r} \frac{r \pi d^3}{6 \eta}.
\end{equation}
Applying the Laplacian, $\laplacian$, to the equation of quasi-static equilibrium  of the circular plate, Eq.~\eqref{eq:LL752}, we get
\begin{equation} \label{eq:lap3zeta}
	D \laplacian^3 \zeta - \laplacian p = 0.
\end{equation}
From~Eq.~\eqref{eq:zetadot} and~\eqref{eq:planpoiseuille}, we have
\begin{equation} \label{eq:lapp}
	\begin{split}
		\laplacian p &= \frac{1}{r} \dod{}{r} r \dod{}{r} p \\
		&= -\frac{1}{r} \dod{}{r} \frac{6 \eta}{\pi d^3} \dot V \\
		&= \frac{12\eta}{d^3} \dot \zeta.
	\end{split}
\end{equation}
When inserting this into the differential equation Eq.~\eqref{eq:lap3zeta}, we get the following 6th order partial differential equation describing the time evolution of $\zeta$,
\begin{equation} \label{eq:lap3zeta2}
	\laplacian^3 \zeta - \frac{12 \eta}{D d^3} \dot{\zeta} = 0.
\end{equation}
We can estimate the time scale for the relaxation of the plate by fluid flow through dimensional analysis.
The parameter ${12 \eta}/{D d^3}$ has the dimension time divided by length to the power of 6.
If we take the characteristic length as the radius, $R$, of the disc, the characteristic flow time must be
\begin{align}
	\begin{split}
		\tau_\text{flow} &= \frac{12 \eta}{D d^3} R^6 \\
		&= 144(1 - \nu^2) \frac{G}{E} \left( \frac{R^6}{h^3 d^3} \right) \tau_\text{M}.
	\end{split}
\end{align}
Here $G$ is the shear modulus of the liquid and $\tau_\text{M} = \eta/G$ the Maxwell relaxation time.
The factor $(1 - \nu^2) G/E$ is of the order of one, whereas
\begin{equation}
	144 \left( \frac{R^6}{h^3 d^3} \right) = 144 \left( 20^6 \right)\approx 10^{10}.
\end{equation}

According to these calculations, at room temperature where $\tau_\text{M} \sim {\SI{1}{\pascal\second}}/{\SI{1}{\giga\pascal}}$ is on the order of \SI{1e-9}{\second}, the flow time will be on the order of 10 seconds ($\tau_\text{flow} \sim 10^{10}\tau_\text{M}$).
Even if this is just an estimate, it shows that when the liquid is cooled and its viscosity increases orders of magnitude, it will start contracting axially and bend the outer discs in the 3-PZ PSG, when the radial flow time exceeds the experimental time scale.

Consequently, the liquid is allowed to continue adjusting its volume (albeit mainly in the axial rather than the radial direction) in the 3-PZ PSG, whereas in the 1-PZ PSG the liquid volume freezes in at some temperature (above $T_g$ during cooling).
This would explain why the typical relaxation times measured in the 1-PZ PSG are faster than those measured in the 3-PZ PSG since the isochoric relaxation times are lower than the isobaric.

To test this conjecture, we added a soft kapton spacer with a thickness of \SI{50}{\micro\metre} between the hub and the sapphire discs to allow for axial contraction of the sample (see Fig.~\ref{fig:PSG_Sapphire}b). Both the 1-PZ (with soft spacers) and the 3-PZ measurement were repeated to ensure identical experimental conditions.
The results are shown in  Fig.~\ref{fig:DC704_Comparison}(c+d).
Again, we see a slight difference in the overall scaling between 1-PZ and 3-PZ measurement, but here the time scales of the spectra are identical.
This is illustrated more clearly in Fig.~\ref{fig:DC704_Comparison}(e) which shows the loss peak frequencies of all four measurements, as well as in Fig.~\ref{fig:DC704_Comparison}(f) showing that the ratio of 1-PZ and 3-PZ loss peak frequencies increases as temperature decreases in the original set of measurements. This is consistent with the interpretation that the thermodynamic boundary condition moves further away from the isobaric case as the temperature is lowered. Ratio of the time scales obtained by 1-PZ and 3-PZ PSG is identically one in the new set of measurements. 

Figure~\ref{fig:DC704_Comparison}(g) shows all spectra from the four measurements scaled on the frequency axis to the loss peak frequency, $f_\text{peak}$, and scaled to the maximum value, $G''_\text{peak}=G''(f_\text{peak})$ on the modulus axis. The plot demonstrates that all spectra have the same shape, irrespective of the thermodynamic boundary conditions. This is consistent with DC704 obeying not only TTS, but even time-temperature-pressure superposition (TTPS)\cite{nielsen_pressure_2008,roed_communication_2013}.

We conclude that the 1-PZ PSG principle is validated as long as the sample is allowed to contract axially.
The outer sapphire discs can for this purpose be considered ``infinitely'' rigid.

Appendix \ref{sec:rubber} contains an implementation of the 1-PZ PSG with steel support that can easily be disassembled to mount solid samples. Pilot measurements on synthetic isoprene rubber (IR) with carbon black N121 filler with this device look promising, but still lack full quantitative agreement with other methods. Further work is ongoing to bring the 1-PZ PSG in better quantitative agreement with other techniques for solid samples.

\section{Summary}
\label{sec:conclusion}
The PSG technique works by measuring the capacitance of PZ disc(s) in a sandwich assembly with thin sample layers.
A mathematical model connects the measured capacitance to the shear modulus of the sample, but isolating $G(\omega)$ from this is not straightforward.
We implemented the Newton-Raphson method for extracting the shear modulus from the measured capacitance spectra and validated our implementation.
The new data analysis approach gives results that are in agreement with results obtained with the previously published method \cite{christensen_rheometer_1995} at low frequencies and performs even better at higher frequencies which approach the resonance of the PSG.
The new method is thus both more precise and extends the frequency range of measurements.
In addition, a correction for sample inertia has been implemented.

We proposed a simplification of the PSG measurement principle using only a single PZ disc instead of three and demonstrated that it is equivalent to the 3-PZ design. 

We showed data for two implementations of the 1-PZ PSG: one based on a fixed assembly with sapphire supports intended for liquid samples and one with steel support for solid samples.

Measurements on liquid were performed with a silicone oil and initially showed a discrepancy in measured time scales between the 1-PZ and 3-PZ PSG, which lead to an analysis of the flow behaviour in the two different cells.
This shows that one needs to be careful to control the boundary conditions.
Liquid samples become very viscous near their glass transition temperature and consequently their flow is extremely slow and depending on the exact details of the measurement cell that can lead to different consequences: in the first implementation of the 1-PZ PSG with thick sapphire supports, the radial flow -- and thus the thermal contraction -- stops when the flow time exceeds the experimental time scale, while for the 3-PZ PSG thermal contraction continues in the axial direction.
The two implementations were brought in quantitative agreement with respect to the material time scales by introducing a soft spacer between the sapphire supports and the hub that allowed the liquid to contract axially. The absolute levels of the shear moduli measured by the two implementations agree to within \SI{10}{\percent}, which is equivalent to measurement-to-measurement variations found in 3-PZ PSG measurements.
We thus consider the 1-PZ PSG technique to be fully validated for liquid samples and ready to deploy as a routine measurement.

The new 1-PZ design has several advantages over the 3-PZ PSG: 1) it is simpler (no need for matching of PZ discs), 2) it reduces the waste of discs when PZ discs break either due to gluing, drilling, or simply because they are brittle, 3) it is more versatile, i.e.~it can be implemented in different ways  and for different sample types.
Here we exemplified this by two different implementations, one with sapphire supports for liquid samples and one with steel supports for solid samples (App.~\ref{sec:rubber}).
In general, it is easier to design a 1-PZ PSG cell for different sample environments, e.g. an oven or a different cryostat system, when there is no requirement of three PZ discs in a stacked configuration electrically isolated from surroundings.
Furthermore, the 1-PZ PSG with sapphire windows allows for simultaneous optical investigations.

The limitations on the modulus resolution remains the same in the 1-PZ PSG as in the 3-PZ PSG, i.e.~the range from \SI{1}{\mega\pascal} to $\sim$\SI{10}{\giga\pascal}, so the method is best suited for relatively stiff samples.

\section*{Acknowledgements}
This work was supported by the VILLUM~Foundation’s \emph{Matter} grant (grant~no.~16515) and by Innovation~Fund~Denmark (case~no.~9065-00002B).
Work at TU~Dortmund was supported by the Deutsche Forschungsgemeinschaft (grant~no.~461147152).

\bibliography{OneDiscPSG}

\begin{thebibliography}{55}%
\makeatletter
\providecommand \@ifxundefined [1]{%
 \@ifx{#1\undefined}
}%
\providecommand \@ifnum [1]{%
 \ifnum #1\expandafter \@firstoftwo
 \else \expandafter \@secondoftwo
 \fi
}%
\providecommand \@ifx [1]{%
 \ifx #1\expandafter \@firstoftwo
 \else \expandafter \@secondoftwo
 \fi
}%
\providecommand \natexlab [1]{#1}%
\providecommand \enquote  [1]{``#1''}%
\providecommand \bibnamefont  [1]{#1}%
\providecommand \bibfnamefont [1]{#1}%
\providecommand \citenamefont [1]{#1}%
\providecommand \href@noop [0]{\@secondoftwo}%
\providecommand \href [0]{\begingroup \@sanitize@url \@href}%
\providecommand \@href[1]{\@@startlink{#1}\@@href}%
\providecommand \@@href[1]{\endgroup#1\@@endlink}%
\providecommand \@sanitize@url [0]{\catcode `\\12\catcode `\$12\catcode
  `\&12\catcode `\#12\catcode `\^12\catcode `\_12\catcode `\%12\relax}%
\providecommand \@@startlink[1]{}%
\providecommand \@@endlink[0]{}%
\providecommand \url  [0]{\begingroup\@sanitize@url \@url }%
\providecommand \@url [1]{\endgroup\@href {#1}{\urlprefix }}%
\providecommand \urlprefix  [0]{URL }%
\providecommand \Eprint [0]{\href }%
\providecommand \doibase [0]{https://doi.org/}%
\providecommand \selectlanguage [0]{\@gobble}%
\providecommand \bibinfo  [0]{\@secondoftwo}%
\providecommand \bibfield  [0]{\@secondoftwo}%
\providecommand \translation [1]{[#1]}%
\providecommand \BibitemOpen [0]{}%
\providecommand \bibitemStop [0]{}%
\providecommand \bibitemNoStop [0]{.\EOS\space}%
\providecommand \EOS [0]{\spacefactor3000\relax}%
\providecommand \BibitemShut  [1]{\csname bibitem#1\endcsname}%
\let\auto@bib@innerbib\@empty
\bibitem [{\citenamefont {Ashby}\ and\ \citenamefont
  {Greer}(2006)}]{ashby_metallic_2006}%
  \BibitemOpen
  \bibfield  {author} {\bibinfo {author} {\bibfnamefont {M.~F.}\ \bibnamefont
  {Ashby}}\ and\ \bibinfo {author} {\bibfnamefont {A.~L.}\ \bibnamefont
  {Greer}},\ }\bibfield  {title} {{\selectlanguage {english}\enquote {\bibinfo
  {title} {Metallic glasses as structural materials},}\ }}\href
  {https://doi.org/10.1016/j.scriptamat.2005.09.051} {\bibfield  {journal}
  {\bibinfo  {journal} {Scripta Materialia}\ }\bibinfo {series} {Viewpoint set
  no: 37. {On} mechanical behavior of metallic glasses},\ \textbf {\bibinfo
  {volume} {54}},\ \bibinfo {pages} {321--326} (\bibinfo {year}
  {2006})}\BibitemShut {NoStop}%
\bibitem [{\citenamefont {Fischer}\ and\ \citenamefont
  {Windhab}(2011)}]{fischer_rheology_2011}%
  \BibitemOpen
  \bibfield  {author} {\bibinfo {author} {\bibfnamefont {P.}~\bibnamefont
  {Fischer}}\ and\ \bibinfo {author} {\bibfnamefont {E.~J.}\ \bibnamefont
  {Windhab}},\ }\bibfield  {title} {{\selectlanguage {english}\enquote
  {\bibinfo {title} {Rheology of food materials},}\ }}\href
  {https://doi.org/10.1016/j.cocis.2010.07.003} {\bibfield  {journal} {\bibinfo
   {journal} {Current Opinion in Colloid \& Interface Science}\ }\textbf
  {\bibinfo {volume} {16}},\ \bibinfo {pages} {36--40} (\bibinfo {year}
  {2011})}\BibitemShut {NoStop}%
\bibitem [{\citenamefont {Angell}(1988)}]{angell_perspective_1988}%
  \BibitemOpen
  \bibfield  {author} {\bibinfo {author} {\bibfnamefont {C.}~\bibnamefont
  {Angell}},\ }\bibfield  {title} {{\selectlanguage {english}\enquote {\bibinfo
  {title} {Perspective on the glass transition},}\ }}\href
  {https://doi.org/10.1016/0022-3697(88)90002-9} {\bibfield  {journal}
  {\bibinfo  {journal} {Journal of Physics and Chemistry of Solids}\ }\textbf
  {\bibinfo {volume} {49}},\ \bibinfo {pages} {863--871} (\bibinfo {year}
  {1988})}\BibitemShut {NoStop}%
\bibitem [{\citenamefont {Mauro}\ \emph {et~al.}(2009)\citenamefont {Mauro},
  \citenamefont {Yue}, \citenamefont {Ellison}, \citenamefont {Gupta},\ and\
  \citenamefont {Allan}}]{mauro_viscosity_2009}%
  \BibitemOpen
  \bibfield  {author} {\bibinfo {author} {\bibfnamefont {J.~C.}\ \bibnamefont
  {Mauro}}, \bibinfo {author} {\bibfnamefont {Y.}~\bibnamefont {Yue}}, \bibinfo
  {author} {\bibfnamefont {A.~J.}\ \bibnamefont {Ellison}}, \bibinfo {author}
  {\bibfnamefont {P.~K.}\ \bibnamefont {Gupta}},\ and\ \bibinfo {author}
  {\bibfnamefont {D.~G.}\ \bibnamefont {Allan}},\ }\bibfield  {title}
  {{\selectlanguage {english}\enquote {\bibinfo {title} {Viscosity of
  glass-forming liquids},}\ }}\href {https://doi.org/10.1073/pnas.0911705106}
  {\bibfield  {journal} {\bibinfo  {journal} {Proceedings of the National
  Academy of Sciences}\ }\textbf {\bibinfo {volume} {106}},\ \bibinfo {pages}
  {19780--19784} (\bibinfo {year} {2009})}\BibitemShut {NoStop}%
\bibitem [{\citenamefont {Hou}\ and\ \citenamefont
  {Kassim}(2005)}]{hou_instrument_2005}%
  \BibitemOpen
  \bibfield  {author} {\bibinfo {author} {\bibfnamefont {Y.~Y.}\ \bibnamefont
  {Hou}}\ and\ \bibinfo {author} {\bibfnamefont {H.~O.}\ \bibnamefont
  {Kassim}},\ }\bibfield  {title} {\enquote {\bibinfo {title} {Instrument
  techniques for rheometry},}\ }\href {https://doi.org/10.1063/1.2085048}
  {\bibfield  {journal} {\bibinfo  {journal} {Review of Scientific
  Instruments}\ }\textbf {\bibinfo {volume} {76}},\ \bibinfo {pages} {101101}
  (\bibinfo {year} {2005})}\BibitemShut {NoStop}%
\bibitem [{\citenamefont {Schroyen}\ \emph {et~al.}(2020)\citenamefont
  {Schroyen}, \citenamefont {Vlassopoulos}, \citenamefont {Van~Puyvelde},\ and\
  \citenamefont {Vermant}}]{schroyen_bulk_2020}%
  \BibitemOpen
  \bibfield  {author} {\bibinfo {author} {\bibfnamefont {B.}~\bibnamefont
  {Schroyen}}, \bibinfo {author} {\bibfnamefont {D.}~\bibnamefont
  {Vlassopoulos}}, \bibinfo {author} {\bibfnamefont {P.}~\bibnamefont
  {Van~Puyvelde}},\ and\ \bibinfo {author} {\bibfnamefont {J.}~\bibnamefont
  {Vermant}},\ }\bibfield  {title} {{\selectlanguage {english}\enquote
  {\bibinfo {title} {Bulk rheometry at high frequencies: a review of
  experimental approaches},}\ }}\href
  {https://doi.org/10.1007/s00397-019-01172-w} {\bibfield  {journal} {\bibinfo
  {journal} {Rheologica Acta}\ }\textbf {\bibinfo {volume} {59}},\ \bibinfo
  {pages} {1--22} (\bibinfo {year} {2020})}\BibitemShut {NoStop}%
\bibitem [{\citenamefont {Athanasiou}\ \emph {et~al.}(2019)\citenamefont
  {Athanasiou}, \citenamefont {Auernhammer}, \citenamefont {Vlassopoulos},\
  and\ \citenamefont {Petekidis}}]{athanasiou_high-frequency_2019}%
  \BibitemOpen
  \bibfield  {author} {\bibinfo {author} {\bibfnamefont {T.}~\bibnamefont
  {Athanasiou}}, \bibinfo {author} {\bibfnamefont {G.~K.}\ \bibnamefont
  {Auernhammer}}, \bibinfo {author} {\bibfnamefont {D.}~\bibnamefont
  {Vlassopoulos}},\ and\ \bibinfo {author} {\bibfnamefont {G.}~\bibnamefont
  {Petekidis}},\ }\bibfield  {title} {{\selectlanguage {english}\enquote
  {\bibinfo {title} {A high-frequency piezoelectric rheometer with validation
  of the loss angle measuring loop: application to polymer melts and colloidal
  glasses},}\ }}\href {https://doi.org/10.1007/s00397-019-01163-x} {\bibfield
  {journal} {\bibinfo  {journal} {Rheologica Acta}\ }\textbf {\bibinfo {volume}
  {58}},\ \bibinfo {pages} {619--637} (\bibinfo {year} {2019})}\BibitemShut
  {NoStop}%
\bibitem [{\citenamefont {Schröter}\ \emph {et~al.}(2006)\citenamefont
  {Schröter}, \citenamefont {Hutcheson}, \citenamefont {Shi}, \citenamefont
  {Mandanici},\ and\ \citenamefont {McKenna}}]{schroter_dynamic_2006}%
  \BibitemOpen
  \bibfield  {author} {\bibinfo {author} {\bibfnamefont {K.}~\bibnamefont
  {Schröter}}, \bibinfo {author} {\bibfnamefont {S.~A.}\ \bibnamefont
  {Hutcheson}}, \bibinfo {author} {\bibfnamefont {X.}~\bibnamefont {Shi}},
  \bibinfo {author} {\bibfnamefont {A.}~\bibnamefont {Mandanici}},\ and\
  \bibinfo {author} {\bibfnamefont {G.~B.}\ \bibnamefont {McKenna}},\
  }\bibfield  {title} {{\selectlanguage {english}\enquote {\bibinfo {title}
  {Dynamic shear modulus of glycerol: {Corrections} due to instrument
  compliance},}\ }}\href {https://doi.org/10.1063/1.2400862} {\bibfield
  {journal} {\bibinfo  {journal} {The Journal of Chemical Physics}\ }\textbf
  {\bibinfo {volume} {125}},\ \bibinfo {pages} {214507} (\bibinfo {year}
  {2006})}\BibitemShut {NoStop}%
\bibitem [{\citenamefont {Laukkanen}(2017)}]{laukkanen_small-diameter_2017}%
  \BibitemOpen
  \bibfield  {author} {\bibinfo {author} {\bibfnamefont {O.-V.}\ \bibnamefont
  {Laukkanen}},\ }\bibfield  {title} {{\selectlanguage {english}\enquote
  {\bibinfo {title} {Small-diameter parallel plate rheometry: a simple
  technique for measuring rheological properties of glass-forming liquids in
  shear},}\ }}\href {https://doi.org/10.1007/s00397-017-1020-5} {\bibfield
  {journal} {\bibinfo  {journal} {Rheologica Acta}\ }\textbf {\bibinfo {volume}
  {56}},\ \bibinfo {pages} {661--671} (\bibinfo {year} {2017})}\BibitemShut
  {NoStop}%
\bibitem [{\citenamefont {Ferry}(1980)}]{ferry_viscoelastic_1980}%
  \BibitemOpen
  \bibfield  {author} {\bibinfo {author} {\bibfnamefont {J.~D.}\ \bibnamefont
  {Ferry}},\ }\href@noop {} {\emph {\bibinfo {title} {Viscoelastic properties
  of polymers}}},\ \bibinfo {edition} {3rd}\ ed.\ (\bibinfo  {publisher} {John
  Wiley \& Sons, Inc},\ \bibinfo {address} {New York},\ \bibinfo {year}
  {1980})\BibitemShut {NoStop}%
\bibitem [{\citenamefont {Sader}(1998)}]{sader_frequency_1998}%
  \BibitemOpen
  \bibfield  {author} {\bibinfo {author} {\bibfnamefont {J.~E.}\ \bibnamefont
  {Sader}},\ }\bibfield  {title} {{\selectlanguage {english}\enquote {\bibinfo
  {title} {Frequency response of cantilever beams immersed in viscous fluids
  with applications to the atomic force microscope},}\ }}\href
  {https://doi.org/10.1063/1.368002} {\bibfield  {journal} {\bibinfo  {journal}
  {Journal of Applied Physics}\ }\textbf {\bibinfo {volume} {84}},\ \bibinfo
  {pages} {64--76} (\bibinfo {year} {1998})}\BibitemShut {NoStop}%
\bibitem [{\citenamefont {Ahmed}, \citenamefont {Nino},\ and\ \citenamefont
  {Moy}(2001)}]{ahmed_measurement_2001}%
  \BibitemOpen
  \bibfield  {author} {\bibinfo {author} {\bibfnamefont {N.}~\bibnamefont
  {Ahmed}}, \bibinfo {author} {\bibfnamefont {D.~F.}\ \bibnamefont {Nino}},\
  and\ \bibinfo {author} {\bibfnamefont {V.~T.}\ \bibnamefont {Moy}},\
  }\bibfield  {title} {{\selectlanguage {english}\enquote {\bibinfo {title}
  {Measurement of solution viscosity by atomic force microscopy},}\ }}\href
  {https://doi.org/10.1063/1.1368856} {\bibfield  {journal} {\bibinfo
  {journal} {Review of Scientific Instruments}\ }\textbf {\bibinfo {volume}
  {72}},\ \bibinfo {pages} {2731--2734} (\bibinfo {year} {2001})}\BibitemShut
  {NoStop}%
\bibitem [{\citenamefont {Christopher}\ \emph {et~al.}(2010)\citenamefont
  {Christopher}, \citenamefont {Yoo}, \citenamefont {Dagalakis}, \citenamefont
  {Hudson},\ and\ \citenamefont {Migler}}]{christopher_development_2010}%
  \BibitemOpen
  \bibfield  {author} {\bibinfo {author} {\bibfnamefont {G.~F.}\ \bibnamefont
  {Christopher}}, \bibinfo {author} {\bibfnamefont {J.~M.}\ \bibnamefont
  {Yoo}}, \bibinfo {author} {\bibfnamefont {N.}~\bibnamefont {Dagalakis}},
  \bibinfo {author} {\bibfnamefont {S.~D.}\ \bibnamefont {Hudson}},\ and\
  \bibinfo {author} {\bibfnamefont {K.~B.}\ \bibnamefont {Migler}},\ }\bibfield
   {title} {{\selectlanguage {english}\enquote {\bibinfo {title} {Development
  of a {MEMS} based dynamic rheometer},}\ }}\href
  {https://doi.org/10.1039/c005065b} {\bibfield  {journal} {\bibinfo  {journal}
  {Lab on a Chip}\ }\textbf {\bibinfo {volume} {10}},\ \bibinfo {pages} {2749}
  (\bibinfo {year} {2010})}\BibitemShut {NoStop}%
\bibitem [{\citenamefont {Mather}\ \emph {et~al.}(2012)\citenamefont {Mather},
  \citenamefont {Rides}, \citenamefont {Allen},\ and\ \citenamefont
  {Tomlins}}]{mather_liquid_2012}%
  \BibitemOpen
  \bibfield  {author} {\bibinfo {author} {\bibfnamefont {M.~L.}\ \bibnamefont
  {Mather}}, \bibinfo {author} {\bibfnamefont {M.}~\bibnamefont {Rides}},
  \bibinfo {author} {\bibfnamefont {C.~R.~G.}\ \bibnamefont {Allen}},\ and\
  \bibinfo {author} {\bibfnamefont {P.~E.}\ \bibnamefont {Tomlins}},\
  }\bibfield  {title} {{\selectlanguage {english}\enquote {\bibinfo {title}
  {Liquid viscoelasticity probed by a mesoscale piezoelectric bimorph
  cantilever},}\ }}\href {https://doi.org/10.1122/1.3670732} {\bibfield
  {journal} {\bibinfo  {journal} {Journal of Rheology}\ }\textbf {\bibinfo
  {volume} {56}},\ \bibinfo {pages} {99--112} (\bibinfo {year}
  {2012})}\BibitemShut {NoStop}%
\bibitem [{\citenamefont {McSkimin}(1964)}]{mcskimin_ultrasonic_1964}%
  \BibitemOpen
  \bibfield  {author} {\bibinfo {author} {\bibfnamefont {H.~J.}\ \bibnamefont
  {McSkimin}},\ }\bibfield  {title} {{\selectlanguage {english}\enquote
  {\bibinfo {title} {Ultrasonic {Methods} for {Measuring} the {Mechanical}
  {Properties} of {Liquids} and {Solids}},}\ }}in\ \href
  {https://doi.org/10.1016/B978-1-4832-2857-0.50010-1} {{\selectlanguage
  {english}\emph {\bibinfo {booktitle} {Physical {Acoustics}: {Principles} and
  methods}}}},\ Vol.~\bibinfo {volume} {IA},\ \bibinfo {editor} {edited by\
  \bibinfo {editor} {\bibfnamefont {W.~P.}\ \bibnamefont {Mason}}}\ (\bibinfo
  {publisher} {Academic Press},\ \bibinfo {year} {1964})\ pp.\ \bibinfo {pages}
  {271--334}\BibitemShut {NoStop}%
\bibitem [{\citenamefont {Yan}\ and\ \citenamefont
  {Nelson}(1987)}]{yan_impulsive_1987}%
  \BibitemOpen
  \bibfield  {author} {\bibinfo {author} {\bibfnamefont {Y.}~\bibnamefont
  {Yan}}\ and\ \bibinfo {author} {\bibfnamefont {K.~A.}\ \bibnamefont
  {Nelson}},\ }\bibfield  {title} {\enquote {\bibinfo {title} {Impulsive
  stimulated light scattering. {I}. {General} theory},}\ }\href
  {https://doi.org/10.1063/1.453733} {\bibfield  {journal} {\bibinfo  {journal}
  {The Journal of Chemical Physics}\ }\textbf {\bibinfo {volume} {87}},\
  \bibinfo {pages} {6240--6256} (\bibinfo {year} {1987})}\BibitemShut {NoStop}%
\bibitem [{\citenamefont {Pick}\ \emph {et~al.}(2003)\citenamefont {Pick},
  \citenamefont {Dreyfus}, \citenamefont {Azzimani}, \citenamefont {Taschin},
  \citenamefont {Ricci}, \citenamefont {Torre},\ and\ \citenamefont
  {Franosch}}]{pick_frequency_2003}%
  \BibitemOpen
  \bibfield  {author} {\bibinfo {author} {\bibfnamefont {R.~M.}\ \bibnamefont
  {Pick}}, \bibinfo {author} {\bibfnamefont {C.}~\bibnamefont {Dreyfus}},
  \bibinfo {author} {\bibfnamefont {A.}~\bibnamefont {Azzimani}}, \bibinfo
  {author} {\bibfnamefont {A.}~\bibnamefont {Taschin}}, \bibinfo {author}
  {\bibfnamefont {M.}~\bibnamefont {Ricci}}, \bibinfo {author} {\bibfnamefont
  {R.}~\bibnamefont {Torre}},\ and\ \bibinfo {author} {\bibfnamefont
  {T.}~\bibnamefont {Franosch}},\ }\bibfield  {title} {{\selectlanguage
  {english}\enquote {\bibinfo {title} {Frequency and time resolved light
  scattering on longitudinal phonons in molecular supercooled liquids},}\
  }}\href {https://doi.org/10.1088/0953-8984/15/11/307} {\bibfield  {journal}
  {\bibinfo  {journal} {Journal of Physics: Condensed Matter}\ }\textbf
  {\bibinfo {volume} {15}},\ \bibinfo {pages} {S825--S834} (\bibinfo {year}
  {2003})}\BibitemShut {NoStop}%
\bibitem [{\citenamefont {Klieber}\ \emph {et~al.}(2012)\citenamefont
  {Klieber}, \citenamefont {Pezeril}, \citenamefont {Andrieu},\ and\
  \citenamefont {Nelson}}]{klieber_optical_2012}%
  \BibitemOpen
  \bibfield  {author} {\bibinfo {author} {\bibfnamefont {C.}~\bibnamefont
  {Klieber}}, \bibinfo {author} {\bibfnamefont {T.}~\bibnamefont {Pezeril}},
  \bibinfo {author} {\bibfnamefont {S.}~\bibnamefont {Andrieu}},\ and\ \bibinfo
  {author} {\bibfnamefont {K.~A.}\ \bibnamefont {Nelson}},\ }\bibfield  {title}
  {\enquote {\bibinfo {title} {Optical generation and detection of
  gigahertz-frequency longitudinal and shear acoustic waves in liquids:
  {Theory} and experiment},}\ }\href {https://doi.org/10.1063/1.4730943}
  {\bibfield  {journal} {\bibinfo  {journal} {Journal of Applied Physics}\
  }\textbf {\bibinfo {volume} {112}},\ \bibinfo {pages} {013502} (\bibinfo
  {year} {2012})}\BibitemShut {NoStop}%
\bibitem [{\citenamefont {Morath}\ and\ \citenamefont
  {Maris}(1996)}]{morath_phonon_1996}%
  \BibitemOpen
  \bibfield  {author} {\bibinfo {author} {\bibfnamefont {C.~J.}\ \bibnamefont
  {Morath}}\ and\ \bibinfo {author} {\bibfnamefont {H.~J.}\ \bibnamefont
  {Maris}},\ }\bibfield  {title} {\enquote {\bibinfo {title} {Phonon
  attenuation in amorphous solids studied by picosecond ultrasonics},}\ }\href
  {https://doi.org/10.1103/PhysRevB.54.203} {\bibfield  {journal} {\bibinfo
  {journal} {Physical Review B}\ }\textbf {\bibinfo {volume} {54}},\ \bibinfo
  {pages} {203--213} (\bibinfo {year} {1996})}\BibitemShut {NoStop}%
\bibitem [{\citenamefont {Pezeril}\ \emph {et~al.}(2009)\citenamefont
  {Pezeril}, \citenamefont {Klieber}, \citenamefont {Andrieu},\ and\
  \citenamefont {Nelson}}]{pezeril_optical_2009}%
  \BibitemOpen
  \bibfield  {author} {\bibinfo {author} {\bibfnamefont {T.}~\bibnamefont
  {Pezeril}}, \bibinfo {author} {\bibfnamefont {C.}~\bibnamefont {Klieber}},
  \bibinfo {author} {\bibfnamefont {S.}~\bibnamefont {Andrieu}},\ and\ \bibinfo
  {author} {\bibfnamefont {K.~A.}\ \bibnamefont {Nelson}},\ }\bibfield  {title}
  {\enquote {\bibinfo {title} {Optical {Generation} of {Gigahertz}-{Frequency}
  {Shear} {Acoustic} {Waves} in {Liquid} {Glycerol}},}\ }\href
  {https://doi.org/10.1103/PhysRevLett.102.107402} {\bibfield  {journal}
  {\bibinfo  {journal} {Physical Review Letters}\ }\textbf {\bibinfo {volume}
  {102}},\ \bibinfo {pages} {107402} (\bibinfo {year} {2009})}\BibitemShut
  {NoStop}%
\bibitem [{\citenamefont {Christensen}\ and\ \citenamefont
  {Olsen}(1995)}]{christensen_rheometer_1995}%
  \BibitemOpen
  \bibfield  {author} {\bibinfo {author} {\bibfnamefont {T.}~\bibnamefont
  {Christensen}}\ and\ \bibinfo {author} {\bibfnamefont {N.~B.}\ \bibnamefont
  {Olsen}},\ }\bibfield  {title} {\enquote {\bibinfo {title} {A rheometer for
  the measurement of a high shear modulus covering more than seven decades of
  frequency below 50 {kHz}},}\ }\href {https://doi.org/10.1063/1.1146126}
  {\bibfield  {journal} {\bibinfo  {journal} {Review of Scientific
  Instruments}\ }\textbf {\bibinfo {volume} {66}},\ \bibinfo {pages}
  {5019--5031} (\bibinfo {year} {1995})}\BibitemShut {NoStop}%
\bibitem [{\citenamefont {Dyre}, \citenamefont {Olsen},\ and\ \citenamefont
  {Christensen}(1996)}]{dyre_local_1996}%
  \BibitemOpen
  \bibfield  {author} {\bibinfo {author} {\bibfnamefont {J.~C.}\ \bibnamefont
  {Dyre}}, \bibinfo {author} {\bibfnamefont {N.~B.}\ \bibnamefont {Olsen}},\
  and\ \bibinfo {author} {\bibfnamefont {T.}~\bibnamefont {Christensen}},\
  }\bibfield  {title} {\enquote {\bibinfo {title} {Local elastic expansion
  model for viscous-flow activation energies of glass-forming molecular
  liquids},}\ }\href {https://doi.org/10.1103/PhysRevB.53.2171} {\bibfield
  {journal} {\bibinfo  {journal} {Physical Review B}\ }\textbf {\bibinfo
  {volume} {53}},\ \bibinfo {pages} {2171--2174} (\bibinfo {year}
  {1996})}\BibitemShut {NoStop}%
\bibitem [{\citenamefont {Eliasen}\ \emph {et~al.}(2021)\citenamefont
  {Eliasen}, \citenamefont {Hansen}, \citenamefont {Lundin}, \citenamefont
  {Rauber}, \citenamefont {Hempelmann}, \citenamefont {Christensen},
  \citenamefont {Hecksher}, \citenamefont {Matic}, \citenamefont {Frick},\ and\
  \citenamefont {Niss}}]{eliasen_high-frequency_2021}%
  \BibitemOpen
  \bibfield  {author} {\bibinfo {author} {\bibfnamefont {K.~L.}\ \bibnamefont
  {Eliasen}}, \bibinfo {author} {\bibfnamefont {H.~W.}\ \bibnamefont {Hansen}},
  \bibinfo {author} {\bibfnamefont {F.}~\bibnamefont {Lundin}}, \bibinfo
  {author} {\bibfnamefont {D.}~\bibnamefont {Rauber}}, \bibinfo {author}
  {\bibfnamefont {R.}~\bibnamefont {Hempelmann}}, \bibinfo {author}
  {\bibfnamefont {T.}~\bibnamefont {Christensen}}, \bibinfo {author}
  {\bibfnamefont {T.}~\bibnamefont {Hecksher}}, \bibinfo {author}
  {\bibfnamefont {A.}~\bibnamefont {Matic}}, \bibinfo {author} {\bibfnamefont
  {B.}~\bibnamefont {Frick}},\ and\ \bibinfo {author} {\bibfnamefont
  {K.}~\bibnamefont {Niss}},\ }\bibfield  {title} {{\selectlanguage
  {english}\enquote {\bibinfo {title} {High-frequency dynamics and test of the
  shoving model for the glass-forming ionic liquid {Pyr14}-{TFSI}},}\ }}\href
  {https://doi.org/10.1103/PhysRevMaterials.5.065606} {\bibfield  {journal}
  {\bibinfo  {journal} {Physical Review Materials}\ }\textbf {\bibinfo {volume}
  {5}},\ \bibinfo {pages} {065606} (\bibinfo {year} {2021})}\BibitemShut
  {NoStop}%
\bibitem [{\citenamefont {Gundermann}\ \emph {et~al.}(2011)\citenamefont
  {Gundermann}, \citenamefont {Pedersen}, \citenamefont {Hecksher},
  \citenamefont {Bailey}, \citenamefont {Jakobsen}, \citenamefont
  {Christensen}, \citenamefont {Olsen}, \citenamefont {Schrøder},
  \citenamefont {Fragiadakis}, \citenamefont {Casalini}, \citenamefont
  {Michael~Roland}, \citenamefont {Dyre},\ and\ \citenamefont
  {Niss}}]{gundermann_predicting_2011}%
  \BibitemOpen
  \bibfield  {author} {\bibinfo {author} {\bibfnamefont {D.}~\bibnamefont
  {Gundermann}}, \bibinfo {author} {\bibfnamefont {U.~R.}\ \bibnamefont
  {Pedersen}}, \bibinfo {author} {\bibfnamefont {T.}~\bibnamefont {Hecksher}},
  \bibinfo {author} {\bibfnamefont {N.~P.}\ \bibnamefont {Bailey}}, \bibinfo
  {author} {\bibfnamefont {B.}~\bibnamefont {Jakobsen}}, \bibinfo {author}
  {\bibfnamefont {T.}~\bibnamefont {Christensen}}, \bibinfo {author}
  {\bibfnamefont {N.~B.}\ \bibnamefont {Olsen}}, \bibinfo {author}
  {\bibfnamefont {T.~B.}\ \bibnamefont {Schrøder}}, \bibinfo {author}
  {\bibfnamefont {D.}~\bibnamefont {Fragiadakis}}, \bibinfo {author}
  {\bibfnamefont {R.}~\bibnamefont {Casalini}}, \bibinfo {author}
  {\bibfnamefont {C.}~\bibnamefont {Michael~Roland}}, \bibinfo {author}
  {\bibfnamefont {J.~C.}\ \bibnamefont {Dyre}},\ and\ \bibinfo {author}
  {\bibfnamefont {K.}~\bibnamefont {Niss}},\ }\bibfield  {title}
  {{\selectlanguage {english}\enquote {\bibinfo {title} {Predicting the
  density-scaling exponent of a glass-forming liquid from {Prigogine}–{Defay}
  ratio measurements},}\ }}\href {https://doi.org/10.1038/nphys2031} {\bibfield
   {journal} {\bibinfo  {journal} {Nature Physics}\ }\textbf {\bibinfo {volume}
  {7}},\ \bibinfo {pages} {816--821} (\bibinfo {year} {2011})}\BibitemShut
  {NoStop}%
\bibitem [{\citenamefont {Gainaru}\ \emph {et~al.}(2014)\citenamefont
  {Gainaru}, \citenamefont {Figuli}, \citenamefont {Hecksher}, \citenamefont
  {Jakobsen}, \citenamefont {Dyre}, \citenamefont {Wilhelm},\ and\
  \citenamefont {Böhmer}}]{gainaru_shear-modulus_2014}%
  \BibitemOpen
  \bibfield  {author} {\bibinfo {author} {\bibfnamefont {C.}~\bibnamefont
  {Gainaru}}, \bibinfo {author} {\bibfnamefont {R.}~\bibnamefont {Figuli}},
  \bibinfo {author} {\bibfnamefont {T.}~\bibnamefont {Hecksher}}, \bibinfo
  {author} {\bibfnamefont {B.}~\bibnamefont {Jakobsen}}, \bibinfo {author}
  {\bibfnamefont {J.~C.}\ \bibnamefont {Dyre}}, \bibinfo {author}
  {\bibfnamefont {M.}~\bibnamefont {Wilhelm}},\ and\ \bibinfo {author}
  {\bibfnamefont {R.}~\bibnamefont {Böhmer}},\ }\bibfield  {title}
  {{\selectlanguage {english}\enquote {\bibinfo {title} {Shear-{Modulus}
  {Investigations} of {Monohydroxy} {Alcohols}: {Evidence} for a
  {Short}-{Chain}-{Polymer} {Rheological} {Response}},}\ }}\href
  {https://doi.org/10.1103/PhysRevLett.112.098301} {\bibfield  {journal}
  {\bibinfo  {journal} {Physical Review Letters}\ }\textbf {\bibinfo {volume}
  {112}},\ \bibinfo {pages} {098301} (\bibinfo {year} {2014})}\BibitemShut
  {NoStop}%
\bibitem [{\citenamefont {Hecksher}\ and\ \citenamefont
  {Jakobsen}(2014)}]{hecksher2014_supra}%
  \BibitemOpen
  \bibfield  {author} {\bibinfo {author} {\bibfnamefont {T.}~\bibnamefont
  {Hecksher}}\ and\ \bibinfo {author} {\bibfnamefont {B.}~\bibnamefont
  {Jakobsen}},\ }\bibfield  {title} {\enquote {\bibinfo {title} {Communication:
  Supramolecular structures in monohydroxy alcohols: Insights from
  shear-mechanical studies of a systematic series of octanol structural
  isomers},}\ }\href {https://doi.org/10.1063/1.4895095} {\bibfield  {journal}
  {\bibinfo  {journal} {The Journal of Chemical Physics}\ }\textbf {\bibinfo
  {volume} {141}},\ \bibinfo {pages} {101104} (\bibinfo {year} {2014})},\
  \Eprint {https://arxiv.org/abs/https://doi.org/10.1063/1.4895095}
  {https://doi.org/10.1063/1.4895095} \BibitemShut {NoStop}%
\bibitem [{\citenamefont {Jensen}\ \emph {et~al.}(2018)\citenamefont {Jensen},
  \citenamefont {Gainaru}, \citenamefont {Alba-Simionesco}, \citenamefont
  {Hecksher},\ and\ \citenamefont {Niss}}]{jensen_slow_2018}%
  \BibitemOpen
  \bibfield  {author} {\bibinfo {author} {\bibfnamefont {M.~H.}\ \bibnamefont
  {Jensen}}, \bibinfo {author} {\bibfnamefont {C.}~\bibnamefont {Gainaru}},
  \bibinfo {author} {\bibfnamefont {C.}~\bibnamefont {Alba-Simionesco}},
  \bibinfo {author} {\bibfnamefont {T.}~\bibnamefont {Hecksher}},\ and\
  \bibinfo {author} {\bibfnamefont {K.}~\bibnamefont {Niss}},\ }\bibfield
  {title} {{\selectlanguage {english}\enquote {\bibinfo {title} {Slow
  rheological mode in glycerol and glycerol–water mixtures},}\ }}\href
  {https://doi.org/10.1039/C7CP06482A} {\bibfield  {journal} {\bibinfo
  {journal} {Physical Chemistry Chemical Physics}\ }\textbf {\bibinfo {volume}
  {20}},\ \bibinfo {pages} {1716--1723} (\bibinfo {year} {2018})}\BibitemShut
  {NoStop}%
\bibitem [{\citenamefont {Hecksher}\ \emph {et~al.}(2017)\citenamefont
  {Hecksher}, \citenamefont {Torchinsky}, \citenamefont {Klieber},
  \citenamefont {Johnson}, \citenamefont {Dyre},\ and\ \citenamefont
  {Nelson}}]{hecksher_toward_2017}%
  \BibitemOpen
  \bibfield  {author} {\bibinfo {author} {\bibfnamefont {T.}~\bibnamefont
  {Hecksher}}, \bibinfo {author} {\bibfnamefont {D.~H.}\ \bibnamefont
  {Torchinsky}}, \bibinfo {author} {\bibfnamefont {C.}~\bibnamefont {Klieber}},
  \bibinfo {author} {\bibfnamefont {J.~A.}\ \bibnamefont {Johnson}}, \bibinfo
  {author} {\bibfnamefont {J.~C.}\ \bibnamefont {Dyre}},\ and\ \bibinfo
  {author} {\bibfnamefont {K.~A.}\ \bibnamefont {Nelson}},\ }\bibfield  {title}
  {{\selectlanguage {english}\enquote {\bibinfo {title} {Toward broadband
  mechanical spectroscopy},}\ }}\href {https://doi.org/10.1073/pnas.1707251114}
  {\bibfield  {journal} {\bibinfo  {journal} {Proceedings of the National
  Academy of Sciences}\ }\textbf {\bibinfo {volume} {114}},\ \bibinfo {pages}
  {8710--8715} (\bibinfo {year} {2017})}\BibitemShut {NoStop}%
\bibitem [{\citenamefont {Hecksher}, \citenamefont {Olsen},\ and\ \citenamefont
  {Dyre}(2017{\natexlab{a}})}]{hecksher2017_squalane}%
  \BibitemOpen
  \bibfield  {author} {\bibinfo {author} {\bibfnamefont {T.}~\bibnamefont
  {Hecksher}}, \bibinfo {author} {\bibfnamefont {N.~B.}\ \bibnamefont
  {Olsen}},\ and\ \bibinfo {author} {\bibfnamefont {J.~C.}\ \bibnamefont
  {Dyre}},\ }\bibfield  {title} {\enquote {\bibinfo {title} {Model for the
  alpha and beta shear-mechanical properties of supercooled liquids and its
  comparison to squalane data},}\ }\href {https://doi.org/10.1063/1.4979658}
  {\bibfield  {journal} {\bibinfo  {journal} {Journal of Chemical Physics}\
  }\textbf {\bibinfo {volume} {146}},\ \bibinfo {pages} {154504} (\bibinfo
  {year} {2017}{\natexlab{a}})},\ \Eprint
  {https://arxiv.org/abs/https://doi.org/10.1063/1.4979658}
  {https://doi.org/10.1063/1.4979658} \BibitemShut {NoStop}%
\bibitem [{\citenamefont {Hecksher}, \citenamefont {Olsen},\ and\ \citenamefont
  {Dyre}(2017{\natexlab{b}})}]{hecksher2022_alphamodel}%
  \BibitemOpen
  \bibfield  {author} {\bibinfo {author} {\bibfnamefont {T.}~\bibnamefont
  {Hecksher}}, \bibinfo {author} {\bibfnamefont {N.~B.}\ \bibnamefont
  {Olsen}},\ and\ \bibinfo {author} {\bibfnamefont {J.~C.}\ \bibnamefont
  {Dyre}},\ }\bibfield  {title} {\enquote {\bibinfo {title} {Model for the
  alpha and beta shear-mechanical properties of supercooled liquids and its
  comparison to squalane data},}\ }\href {https://doi.org/10.1063/1.4979658}
  {\bibfield  {journal} {\bibinfo  {journal} {Journal of Chemical Physics}\
  }\textbf {\bibinfo {volume} {146}},\ \bibinfo {pages} {154504} (\bibinfo
  {year} {2017}{\natexlab{b}})},\ \Eprint
  {https://arxiv.org/abs/https://doi.org/10.1063/1.4979658}
  {https://doi.org/10.1063/1.4979658} \BibitemShut {NoStop}%
\bibitem [{\citenamefont {Jakobsen}\ \emph {et~al.}(2012)\citenamefont
  {Jakobsen}, \citenamefont {Hecksher}, \citenamefont {Christensen},
  \citenamefont {Olsen}, \citenamefont {Dyre},\ and\ \citenamefont
  {Niss}}]{jakobsen_communication_2012}%
  \BibitemOpen
  \bibfield  {author} {\bibinfo {author} {\bibfnamefont {B.}~\bibnamefont
  {Jakobsen}}, \bibinfo {author} {\bibfnamefont {T.}~\bibnamefont {Hecksher}},
  \bibinfo {author} {\bibfnamefont {T.}~\bibnamefont {Christensen}}, \bibinfo
  {author} {\bibfnamefont {N.~B.}\ \bibnamefont {Olsen}}, \bibinfo {author}
  {\bibfnamefont {J.~C.}\ \bibnamefont {Dyre}},\ and\ \bibinfo {author}
  {\bibfnamefont {K.}~\bibnamefont {Niss}},\ }\bibfield  {title} {\enquote
  {\bibinfo {title} {Communication: {Identical} temperature dependence of the
  time scales of several linear-response functions of two glass-forming
  liquids},}\ }\href {https://doi.org/10.1063/1.3690083} {\bibfield  {journal}
  {\bibinfo  {journal} {The Journal of Chemical Physics}\ }\textbf {\bibinfo
  {volume} {136}},\ \bibinfo {pages} {081102} (\bibinfo {year}
  {2012})}\BibitemShut {NoStop}%
\bibitem [{\citenamefont {Roed}\ \emph {et~al.}(2021)\citenamefont {Roed},
  \citenamefont {Dyre}, \citenamefont {Niss}, \citenamefont {Hecksher},\ and\
  \citenamefont {Riechers}}]{roed_time-scale_2021}%
  \BibitemOpen
  \bibfield  {author} {\bibinfo {author} {\bibfnamefont {L.~A.}\ \bibnamefont
  {Roed}}, \bibinfo {author} {\bibfnamefont {J.~C.}\ \bibnamefont {Dyre}},
  \bibinfo {author} {\bibfnamefont {K.}~\bibnamefont {Niss}}, \bibinfo {author}
  {\bibfnamefont {T.}~\bibnamefont {Hecksher}},\ and\ \bibinfo {author}
  {\bibfnamefont {B.}~\bibnamefont {Riechers}},\ }\bibfield  {title}
  {{\selectlanguage {english}\enquote {\bibinfo {title} {Time-scale ordering in
  hydrogen- and van der {Waals}-bonded liquids},}\ }}\href
  {https://doi.org/10.1063/5.0049108} {\bibfield  {journal} {\bibinfo
  {journal} {The Journal of Chemical Physics}\ }\textbf {\bibinfo {volume}
  {154}},\ \bibinfo {pages} {184508} (\bibinfo {year} {2021})}\BibitemShut
  {NoStop}%
\bibitem [{\citenamefont {Weigl}\ \emph {et~al.}(2021)\citenamefont {Weigl},
  \citenamefont {Hecksher}, \citenamefont {Dyre}, \citenamefont {Walther},\
  and\ \citenamefont {Blochowicz}}]{weigl_identity_2021}%
  \BibitemOpen
  \bibfield  {author} {\bibinfo {author} {\bibfnamefont {P.}~\bibnamefont
  {Weigl}}, \bibinfo {author} {\bibfnamefont {T.}~\bibnamefont {Hecksher}},
  \bibinfo {author} {\bibfnamefont {J.~C.}\ \bibnamefont {Dyre}}, \bibinfo
  {author} {\bibfnamefont {T.}~\bibnamefont {Walther}},\ and\ \bibinfo {author}
  {\bibfnamefont {T.}~\bibnamefont {Blochowicz}},\ }\bibfield  {title}
  {{\selectlanguage {english}\enquote {\bibinfo {title} {Identity of the local
  and macroscopic dynamic elastic responses in supercooled 1-propanol},}\
  }}\href {https://doi.org/10.1039/D1CP02671B} {\bibfield  {journal} {\bibinfo
  {journal} {Physical Chemistry Chemical Physics}\ } (\bibinfo {year} {2021}),\
  10.1039/D1CP02671B}\BibitemShut {NoStop}%
\bibitem [{\citenamefont {Christensen}\ and\ \citenamefont
  {Olsen}(1994)}]{christensen_determination_1994}%
  \BibitemOpen
  \bibfield  {author} {\bibinfo {author} {\bibfnamefont {T.}~\bibnamefont
  {Christensen}}\ and\ \bibinfo {author} {\bibfnamefont {N.~B.}\ \bibnamefont
  {Olsen}},\ }\bibfield  {title} {\enquote {\bibinfo {title} {Determination of
  the frequency-dependent bulk modulus of glycerol using a piezoelectric
  spherical shell},}\ }\href {https://doi.org/10.1103/PhysRevB.49.15396}
  {\bibfield  {journal} {\bibinfo  {journal} {Physical Review B}\ }\textbf
  {\bibinfo {volume} {49}},\ \bibinfo {pages} {15396--15399} (\bibinfo {year}
  {1994})}\BibitemShut {NoStop}%
\bibitem [{\citenamefont {{Meggitt A/S}}(2019)}]{meggitt_as_data_2019}%
  \BibitemOpen
  \bibfield  {author} {\bibinfo {author} {\bibnamefont {{Meggitt A/S}}},\
  }\href
  {https://www.meggittferroperm.com/wp-content/uploads/2021/10/Datasheet-hard-pz26.pdf}
  {{\selectlanguage {english}\enquote {\bibinfo {title} {Data {Sheet}: {Hard}
  relaxor type {PZT} {Type} {Pz26} ({Navy} {I})},}\ }} (\bibinfo {year}
  {2019}),\ \bibinfo {note}
  {https://www.meggittferroperm.com/resources/data-sheets/}\BibitemShut
  {NoStop}%
\bibitem [{\citenamefont {Schrag}(1977)}]{schrag_deviation_1977}%
  \BibitemOpen
  \bibfield  {author} {\bibinfo {author} {\bibfnamefont {J.~L.}\ \bibnamefont
  {Schrag}},\ }\bibfield  {title} {\enquote {\bibinfo {title} {Deviation of
  {Velocity} {Gradient} {Profiles} from the “{Gap} {Loading}” and
  “{Surface} {Loading}” {Limits} in {Dynamic} {Simple} {Shear}
  {Experiments}},}\ }\href {https://doi.org/10.1122/1.549445} {\bibfield
  {journal} {\bibinfo  {journal} {Transactions of the Society of Rheology}\
  }\textbf {\bibinfo {volume} {21}},\ \bibinfo {pages} {399--413} (\bibinfo
  {year} {1977})},\ \bibinfo {note} {publisher: The Society of
  Rheology}\BibitemShut {NoStop}%
\bibitem [{Note1()}]{Note1}%
  \BibitemOpen
  \bibinfo {note} {In \protect \citeauthor {christensen_rheometer_1995} the
  symbol $p$ was used in stead of $\nu $. We change it here to be consistent
  throughout the current paper.}\BibitemShut {Stop}%
\bibitem [{\citenamefont {Ryaben'kii}\ and\ \citenamefont
  {Tsynkov}(2006)}]{Ryabenkii2006}%
  \BibitemOpen
  \bibfield  {author} {\bibinfo {author} {\bibfnamefont {V.~S.}\ \bibnamefont
  {Ryaben'kii}}\ and\ \bibinfo {author} {\bibfnamefont {S.~V.}\ \bibnamefont
  {Tsynkov}},\ }\href@noop {} {\emph {\bibinfo {title} {A Theoretical
  Introduction to Numerical Analysis}}}\ (\bibinfo  {publisher} {Taylor \&
  Francis Inc},\ \bibinfo {year} {2006})\BibitemShut {NoStop}%
\bibitem [{\citenamefont {Schröter}\ and\ \citenamefont
  {Donth}(2000)}]{schroter_viscosity_2000}%
  \BibitemOpen
  \bibfield  {author} {\bibinfo {author} {\bibfnamefont {K.}~\bibnamefont
  {Schröter}}\ and\ \bibinfo {author} {\bibfnamefont {E.}~\bibnamefont
  {Donth}},\ }\bibfield  {title} {{\selectlanguage {english}\enquote {\bibinfo
  {title} {Viscosity and shear response at the dynamic glass transition of
  glycerol},}\ }}\href {https://doi.org/10.1063/1.1319616} {\bibfield
  {journal} {\bibinfo  {journal} {The Journal of Chemical Physics}\ }\textbf
  {\bibinfo {volume} {113}},\ \bibinfo {pages} {9101--9108} (\bibinfo {year}
  {2000})}\BibitemShut {NoStop}%
\bibitem [{\citenamefont {Scarponi}\ \emph {et~al.}(2004)\citenamefont
  {Scarponi}, \citenamefont {Comez}, \citenamefont {Fioretto},\ and\
  \citenamefont {Palmieri}}]{scarponi_brillouin_2004}%
  \BibitemOpen
  \bibfield  {author} {\bibinfo {author} {\bibfnamefont {F.}~\bibnamefont
  {Scarponi}}, \bibinfo {author} {\bibfnamefont {L.}~\bibnamefont {Comez}},
  \bibinfo {author} {\bibfnamefont {D.}~\bibnamefont {Fioretto}},\ and\
  \bibinfo {author} {\bibfnamefont {L.}~\bibnamefont {Palmieri}},\ }\bibfield
  {title} {{\selectlanguage {english}\enquote {\bibinfo {title} {Brillouin
  light scattering from transverse and longitudinal acoustic waves in
  glycerol},}\ }}\href {https://doi.org/10.1103/PhysRevB.70.054203} {\bibfield
  {journal} {\bibinfo  {journal} {Physical Review B}\ }\textbf {\bibinfo
  {volume} {70}},\ \bibinfo {pages} {054203} (\bibinfo {year}
  {2004})}\BibitemShut {NoStop}%
\bibitem [{\citenamefont {Klieber}\ \emph {et~al.}(2013)\citenamefont
  {Klieber}, \citenamefont {Hecksher}, \citenamefont {Pezeril}, \citenamefont
  {Torchinsky}, \citenamefont {Dyre},\ and\ \citenamefont
  {Nelson}}]{klieber_mechanical_2013}%
  \BibitemOpen
  \bibfield  {author} {\bibinfo {author} {\bibfnamefont {C.}~\bibnamefont
  {Klieber}}, \bibinfo {author} {\bibfnamefont {T.}~\bibnamefont {Hecksher}},
  \bibinfo {author} {\bibfnamefont {T.}~\bibnamefont {Pezeril}}, \bibinfo
  {author} {\bibfnamefont {D.~H.}\ \bibnamefont {Torchinsky}}, \bibinfo
  {author} {\bibfnamefont {J.~C.}\ \bibnamefont {Dyre}},\ and\ \bibinfo
  {author} {\bibfnamefont {K.~A.}\ \bibnamefont {Nelson}},\ }\bibfield  {title}
  {{\selectlanguage {english}\enquote {\bibinfo {title} {Mechanical spectra of
  glass-forming liquids. {II}. {Gigahertz}-frequency longitudinal and shear
  acoustic dynamics in glycerol and {DC704} studied by time-domain {Brillouin}
  scattering},}\ }}\href {https://doi.org/10.1063/1.4789948} {\bibfield
  {journal} {\bibinfo  {journal} {The Journal of Chemical Physics}\ }\textbf
  {\bibinfo {volume} {138}},\ \bibinfo {pages} {12A544} (\bibinfo {year}
  {2013})}\BibitemShut {NoStop}%
\bibitem [{\citenamefont {Igarashi}\ \emph
  {et~al.}(2008{\natexlab{a}})\citenamefont {Igarashi}, \citenamefont
  {Christensen}, \citenamefont {Larsen}, \citenamefont {Olsen}, \citenamefont
  {Pedersen}, \citenamefont {Rasmussen},\ and\ \citenamefont
  {Dyre}}]{igarashi_cryostat_2008}%
  \BibitemOpen
  \bibfield  {author} {\bibinfo {author} {\bibfnamefont {B.}~\bibnamefont
  {Igarashi}}, \bibinfo {author} {\bibfnamefont {T.}~\bibnamefont
  {Christensen}}, \bibinfo {author} {\bibfnamefont {E.~H.}\ \bibnamefont
  {Larsen}}, \bibinfo {author} {\bibfnamefont {N.~B.}\ \bibnamefont {Olsen}},
  \bibinfo {author} {\bibfnamefont {I.~H.}\ \bibnamefont {Pedersen}}, \bibinfo
  {author} {\bibfnamefont {T.}~\bibnamefont {Rasmussen}},\ and\ \bibinfo
  {author} {\bibfnamefont {J.~C.}\ \bibnamefont {Dyre}},\ }\bibfield  {title}
  {{\selectlanguage {english}\enquote {\bibinfo {title} {A cryostat and
  temperature control system optimized for measuring relaxations of
  glass-forming liquids},}\ }}\href {https://doi.org/10.1063/1.2903419}
  {\bibfield  {journal} {\bibinfo  {journal} {Review of Scientific
  Instruments}\ }\textbf {\bibinfo {volume} {79}},\ \bibinfo {pages} {045105}
  (\bibinfo {year} {2008}{\natexlab{a}})}\BibitemShut {NoStop}%
\bibitem [{\citenamefont {Igarashi}\ \emph
  {et~al.}(2008{\natexlab{b}})\citenamefont {Igarashi}, \citenamefont
  {Christensen}, \citenamefont {Larsen}, \citenamefont {Olsen}, \citenamefont
  {Pedersen}, \citenamefont {Rasmussen},\ and\ \citenamefont
  {Dyre}}]{igarashi_impedance-measurement_2008}%
  \BibitemOpen
  \bibfield  {author} {\bibinfo {author} {\bibfnamefont {B.}~\bibnamefont
  {Igarashi}}, \bibinfo {author} {\bibfnamefont {T.}~\bibnamefont
  {Christensen}}, \bibinfo {author} {\bibfnamefont {E.~H.}\ \bibnamefont
  {Larsen}}, \bibinfo {author} {\bibfnamefont {N.~B.}\ \bibnamefont {Olsen}},
  \bibinfo {author} {\bibfnamefont {I.~H.}\ \bibnamefont {Pedersen}}, \bibinfo
  {author} {\bibfnamefont {T.}~\bibnamefont {Rasmussen}},\ and\ \bibinfo
  {author} {\bibfnamefont {J.~C.}\ \bibnamefont {Dyre}},\ }\bibfield  {title}
  {\enquote {\bibinfo {title} {An impedance-measurement setup optimized for
  measuring relaxations of glass-forming liquids},}\ }\href
  {https://doi.org/10.1063/1.2906401} {\bibfield  {journal} {\bibinfo
  {journal} {Review of Scientific Instruments}\ }\textbf {\bibinfo {volume}
  {79}},\ \bibinfo {pages} {045106} (\bibinfo {year}
  {2008}{\natexlab{b}})}\BibitemShut {NoStop}%
\bibitem [{\citenamefont {Niss}\ and\ \citenamefont
  {Hecksher}(2018)}]{niss_perspective:_2018}%
  \BibitemOpen
  \bibfield  {author} {\bibinfo {author} {\bibfnamefont {K.}~\bibnamefont
  {Niss}}\ and\ \bibinfo {author} {\bibfnamefont {T.}~\bibnamefont
  {Hecksher}},\ }\bibfield  {title} {{\selectlanguage {english}\enquote
  {\bibinfo {title} {Perspective: {Searching} for simplicity rather than
  universality in glass-forming liquids},}\ }}\href
  {https://doi.org/10.1063/1.5048093} {\bibfield  {journal} {\bibinfo
  {journal} {The Journal of Chemical Physics}\ }\textbf {\bibinfo {volume}
  {149}},\ \bibinfo {pages} {230901} (\bibinfo {year} {2018})}\BibitemShut
  {NoStop}%
\bibitem [{\citenamefont {Niss}, \citenamefont {Jakobsen},\ and\ \citenamefont
  {Olsen}(2005)}]{niss_dielectric_2005}%
  \BibitemOpen
  \bibfield  {author} {\bibinfo {author} {\bibfnamefont {K.}~\bibnamefont
  {Niss}}, \bibinfo {author} {\bibfnamefont {B.}~\bibnamefont {Jakobsen}},\
  and\ \bibinfo {author} {\bibfnamefont {N.~B.}\ \bibnamefont {Olsen}},\
  }\bibfield  {title} {{\selectlanguage {english}\enquote {\bibinfo {title}
  {Dielectric and shear mechanical relaxations in glass-forming liquids: {A}
  test of the {Gemant}-{DiMarzio}-{Bishop} model},}\ }}\href
  {https://doi.org/10.1063/1.2136886} {\bibfield  {journal} {\bibinfo
  {journal} {The Journal of Chemical Physics}\ }\textbf {\bibinfo {volume}
  {123}},\ \bibinfo {pages} {234510} (\bibinfo {year} {2005})}\BibitemShut
  {NoStop}%
\bibitem [{\citenamefont {Jakobsen}, \citenamefont {Niss},\ and\ \citenamefont
  {Olsen}(2005)}]{jakobsen_dielectric_2005}%
  \BibitemOpen
  \bibfield  {author} {\bibinfo {author} {\bibfnamefont {B.}~\bibnamefont
  {Jakobsen}}, \bibinfo {author} {\bibfnamefont {K.}~\bibnamefont {Niss}},\
  and\ \bibinfo {author} {\bibfnamefont {N.~B.}\ \bibnamefont {Olsen}},\
  }\bibfield  {title} {{\selectlanguage {english}\enquote {\bibinfo {title}
  {Dielectric and shear mechanical alpha and beta relaxations in seven
  glass-forming liquids},}\ }}\href {https://doi.org/10.1063/1.2136887}
  {\bibfield  {journal} {\bibinfo  {journal} {The Journal of Chemical Physics}\
  }\textbf {\bibinfo {volume} {123}},\ \bibinfo {pages} {234511} (\bibinfo
  {year} {2005})}\BibitemShut {NoStop}%
\bibitem [{\citenamefont {Hecksher}(2011)}]{hecksher_relaxation_2011}%
  \BibitemOpen
  \bibfield  {author} {\bibinfo {author} {\bibfnamefont {T.}~\bibnamefont
  {Hecksher}},\ }{\selectlanguage {english}\emph {\bibinfo {title} {Relaxation
  in {Supercooled} {Liquids}}}},\ \href@noop {} {\bibinfo {type} {{PhD}
  thesis}},\ \bibinfo  {school} {Roskilde University}, \bibinfo {address}
  {Roskilde} (\bibinfo {year} {2011})\BibitemShut {NoStop}%
\bibitem [{\citenamefont {Hecksher}\ \emph {et~al.}(2013)\citenamefont
  {Hecksher}, \citenamefont {Olsen}, \citenamefont {Nelson}, \citenamefont
  {Dyre},\ and\ \citenamefont {Christensen}}]{hecksher_mechanical_2013}%
  \BibitemOpen
  \bibfield  {author} {\bibinfo {author} {\bibfnamefont {T.}~\bibnamefont
  {Hecksher}}, \bibinfo {author} {\bibfnamefont {N.~B.}\ \bibnamefont {Olsen}},
  \bibinfo {author} {\bibfnamefont {K.~A.}\ \bibnamefont {Nelson}}, \bibinfo
  {author} {\bibfnamefont {J.~C.}\ \bibnamefont {Dyre}},\ and\ \bibinfo
  {author} {\bibfnamefont {T.}~\bibnamefont {Christensen}},\ }\bibfield
  {title} {{\selectlanguage {english}\enquote {\bibinfo {title} {Mechanical
  spectra of glass-forming liquids. {I}. {Low}-frequency bulk and shear moduli
  of {DC704} and 5-{PPE} measured by piezoceramic transducers},}\ }}\href
  {https://doi.org/10.1063/1.4789946} {\bibfield  {journal} {\bibinfo
  {journal} {The Journal of Chemical Physics}\ }\textbf {\bibinfo {volume}
  {138}},\ \bibinfo {pages} {12A543} (\bibinfo {year} {2013})}\BibitemShut
  {NoStop}%
\bibitem [{\citenamefont {Niss}\ \emph {et~al.}(2012)\citenamefont {Niss},
  \citenamefont {Gundermann}, \citenamefont {Christensen},\ and\ \citenamefont
  {Dyre}}]{niss_dynamic_2012}%
  \BibitemOpen
  \bibfield  {author} {\bibinfo {author} {\bibfnamefont {K.}~\bibnamefont
  {Niss}}, \bibinfo {author} {\bibfnamefont {D.}~\bibnamefont {Gundermann}},
  \bibinfo {author} {\bibfnamefont {T.}~\bibnamefont {Christensen}},\ and\
  \bibinfo {author} {\bibfnamefont {J.~C.}\ \bibnamefont {Dyre}},\ }\bibfield
  {title} {{\selectlanguage {english}\enquote {\bibinfo {title} {Dynamic
  thermal expansivity of liquids near the glass transition},}\ }}\href
  {https://doi.org/10.1103/PhysRevE.85.041501} {\bibfield  {journal} {\bibinfo
  {journal} {Physical Review E}\ }\textbf {\bibinfo {volume} {85}},\ \bibinfo
  {pages} {041501} (\bibinfo {year} {2012})}\BibitemShut {NoStop}%
\bibitem [{\citenamefont {Landau}\ and\ \citenamefont
  {Lifshitz}(1986)}]{landau_theory_1986}%
  \BibitemOpen
  \bibfield  {author} {\bibinfo {author} {\bibfnamefont {L.~D.}\ \bibnamefont
  {Landau}}\ and\ \bibinfo {author} {\bibfnamefont {E.~M.}\ \bibnamefont
  {Lifshitz}},\ }\href@noop {} {\emph {\bibinfo {title} {Theory of
  {Elasticity}}}}\ (\bibinfo  {publisher} {Elsevier},\ \bibinfo {year}
  {1986})\BibitemShut {NoStop}%
\bibitem [{\citenamefont {Nielsen}\ \emph {et~al.}(2008)\citenamefont
  {Nielsen}, \citenamefont {Pawlus}, \citenamefont {Paluch},\ and\
  \citenamefont {Dyre}}]{nielsen_pressure_2008}%
  \BibitemOpen
  \bibfield  {author} {\bibinfo {author} {\bibfnamefont {A.~I.}\ \bibnamefont
  {Nielsen}}, \bibinfo {author} {\bibfnamefont {S.}~\bibnamefont {Pawlus}},
  \bibinfo {author} {\bibfnamefont {M.}~\bibnamefont {Paluch}},\ and\ \bibinfo
  {author} {\bibfnamefont {J.~C.}\ \bibnamefont {Dyre}},\ }\bibfield  {title}
  {{\selectlanguage {english}\enquote {\bibinfo {title} {Pressure dependence of
  the dielectric loss minimum slope for ten molecular liquids},}\ }}\href
  {https://doi.org/10.1080/14786430802607093} {\bibfield  {journal} {\bibinfo
  {journal} {Philosophical Magazine}\ }\textbf {\bibinfo {volume} {88}},\
  \bibinfo {pages} {4101--4108} (\bibinfo {year} {2008})}\BibitemShut {NoStop}%
\bibitem [{\citenamefont {Roed}\ \emph {et~al.}(2013)\citenamefont {Roed},
  \citenamefont {Gundermann}, \citenamefont {Dyre},\ and\ \citenamefont
  {Niss}}]{roed_communication_2013}%
  \BibitemOpen
  \bibfield  {author} {\bibinfo {author} {\bibfnamefont {L.~A.}\ \bibnamefont
  {Roed}}, \bibinfo {author} {\bibfnamefont {D.}~\bibnamefont {Gundermann}},
  \bibinfo {author} {\bibfnamefont {J.~C.}\ \bibnamefont {Dyre}},\ and\
  \bibinfo {author} {\bibfnamefont {K.}~\bibnamefont {Niss}},\ }\bibfield
  {title} {{\selectlanguage {english}\enquote {\bibinfo {title} {Communication:
  {Two} measures of isochronal superposition},}\ }}\href
  {https://doi.org/10.1063/1.4821163} {\bibfield  {journal} {\bibinfo
  {journal} {The Journal of Chemical Physics}\ }\textbf {\bibinfo {volume}
  {139}},\ \bibinfo {pages} {101101} (\bibinfo {year} {2013})}\BibitemShut
  {NoStop}%
\bibitem [{\citenamefont {{ASTM
  International}}(2021{\natexlab{a}})}]{astm_international_astm_2021}%
  \BibitemOpen
  \bibfield  {author} {\bibinfo {author} {\bibnamefont {{ASTM
  International}}},\ }\bibfield  {title} {{\selectlanguage {english}\enquote
  {\bibinfo {title} {{ASTM} {D3191}-10(2020), {Standard} {Test} {Methods} for
  {Carbon} {Black} in {SBR} ({Styrene}-{Butadiene} {Rubber}) — {Recipe} and
  {Evaluation} {Procedures}},}\ }}in\ \href
  {https://www.astm.org/d3191-10r20.html} {{\selectlanguage {english}\emph
  {\bibinfo {booktitle} {Annual {Book} of {ASTM} {Standards}}}}},\ Vol.\
  \bibinfo {volume} {09.01}\ (\bibinfo  {publisher} {ASTM International},\
  \bibinfo {address} {West Conshohocken},\ \bibinfo {year} {2021})\ \bibinfo
  {note} {https://www.astm.org/d3191-10r20.html}\BibitemShut {NoStop}%
\bibitem [{\citenamefont {{ASTM
  International}}(2021{\natexlab{b}})}]{astm_international_astm_2021-1}%
  \BibitemOpen
  \bibfield  {author} {\bibinfo {author} {\bibnamefont {{ASTM
  International}}},\ }\bibfield  {title} {{\selectlanguage {english}\enquote
  {\bibinfo {title} {{ASTM} {D1765}–21 {Standard} {Classification} {System}
  for {Carbon} {Blacks} {Used} in {Rubber} {Products}},}\ }}in\ \href
  {https://www.astm.org/d1765-21.html} {{\selectlanguage {english}\emph
  {\bibinfo {booktitle} {Annual {Book} of {ASTM} {Standards}}}}},\ Vol.\
  \bibinfo {volume} {09.01}\ (\bibinfo  {publisher} {ASTM International},\
  \bibinfo {address} {West Conshohocken},\ \bibinfo {year} {2021})\ \bibinfo
  {note} {https://www.astm.org/d1765-21.html}\BibitemShut {NoStop}%
\bibitem [{\citenamefont {Williams}, \citenamefont {Landel},\ and\
  \citenamefont {Ferry}(1955)}]{williams_temperature_1955}%
  \BibitemOpen
  \bibfield  {author} {\bibinfo {author} {\bibfnamefont {M.~L.}\ \bibnamefont
  {Williams}}, \bibinfo {author} {\bibfnamefont {R.~F.}\ \bibnamefont
  {Landel}},\ and\ \bibinfo {author} {\bibfnamefont {J.~D.}\ \bibnamefont
  {Ferry}},\ }\bibfield  {title} {\enquote {\bibinfo {title} {The {Temperature}
  {Dependence} of {Relaxation} {Mechanisms} in {Amorphous} {Polymers} and
  {Other} {Glass}-forming {Liquids}},}\ }\href
  {https://doi.org/10.1021/ja01619a008} {\bibfield  {journal} {\bibinfo
  {journal} {Journal of the American Chemical Society}\ }\textbf {\bibinfo
  {volume} {77}},\ \bibinfo {pages} {3701--3707} (\bibinfo {year}
  {1955})}\BibitemShut {NoStop}%
\end{thebibliography}%

\appendix
\section{Mathematical derivation of the mapping from 3-PZ PSG to the one-one configuration}
\label{sec:mapping}

For the 3-PZ PSG numerate the top piezo-disc by $1$, the middle one by $2$ and the bottom one by $3$.
The equations of motion for $1$ and $3$ are identical.
By $f_1$ we denote the stress from the liquid layer on the upper disc.
$h$ is the thickness of the disc, $\rho$ its density and $u_1$ the radial displacement.
Newtons $2$nd law for a volume element becomes
\begin{equation}
	\begin{split}
		&(r + \dif r) \dif \phi h \sigma^{(1)}_{r + \dif r,r + \dif r} - r \dif \phi h \sigma^{(1)}_{rr} \\
		&\quad - \dif rh \dif \phi \sigma^{(1)}_{\phi,\phi} + f_1 \dif r r \dif \phi \\
		&= \rho h \dif r r \dif \phi \ddot u_1,
	\end{split}
\end{equation}
and dividing by $\dif r \dif \phi h$ gives for plate $1$
\begin{equation}\label{eq:eqm1}
	\dpd{}{r} r \sigma^{(1)}_{rr} - \sigma^{(1)}_{\phi,\phi} + f_1 \frac{r}{h} = \rho r \ddot u_1,
\end{equation}
which is Eq.~(18) of \citet{christensen_rheometer_1995} (with $\sigma_l = - f_1$).
The stresses can be expressed in terms of the displacement fields using the constitutive equations and expressions of strains (Eqs.~(8) and~(2) of \citet{christensen_rheometer_1995}).
Using the fact that $E_z$ is independent of $r$, this leads to the substitution rule
\begin{equation}
	\dpd{}{r} (r \sigma_{rr}) - \sigma_{\phi,\phi} \rightarrow c_{11} \left( ru'' + u' - \frac{u}{r} \right),
\end{equation}
by which Eq.~\eqref{eq:eqm1} becomes
\begin{equation}\label{eq:p1}
	c_{11} \left( r u_1'' + u_1' - \frac{u_1}{r} \right)
	+ f_1 \frac{r}{h} = \rho r \ddot u_1.
\end{equation}
Correspondingly, we have for plate $2$
\begin{equation}\label{eq:p2}
	c_{11} \left( ru_2'' + u_2' - \frac{u_2}{r} \right) + 2 f_2 \frac{r}{h} = \rho r \ddot u_2,
\end{equation}
where $f_2$ is the stress from one of the liquid layers.
For plate $3$ we get the same equation as for plate $1$,
\begin{equation}\label{eq:p3}
	c_{11} \left( ru_3'' + u_3' - \frac{u_3}{r} \right) + f_3 \frac{r}{h} = \rho r \ddot u_3.
\end{equation}
\begin{figure}
	\centering
	\includegraphics{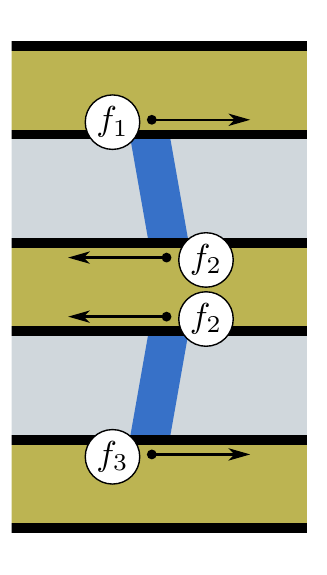}
	\caption{
		Stresses from sample on discs in the 3-PZ PSG.
		This figure shows a section of the discs with sample between the discs (cf. also Fig.~\ref{fig:PSG_Schematic}).
		One liquid filament (shaded in blue) exerts stress on two discs.
		The top liquid filament exerts stress $f_1$ on the upper disc and stress $f_2$ on the middle disc.
		The bottom liquid filament exerts $f_2$ on the middle disc and $f_3$ on the bottom disc.
		The middle disc is thus under a shear stress of $2 f_2$.
	}
	\label{fig:PSG_ForceBalance}
\end{figure}

Assuming a (radially dependent) shear deformation of the two liquid layers we have by symmetry and Newtons $3$rd law that
\begin{equation}\label{eq:shearforces}
	f_2 = -f_1 = -f_3 = -G \frac{u_2 - u_1}{d} = -G \frac{u_2 - u_3}{d}.
\end{equation}
We assume that the time dependence of the fields is proportional to $e^{-i \omega t}$ and redefine $u$ to be the complex amplitude.
This corresponds to the substitution $\ddot u \rightarrow  - \omega^{2} u$.
When inserting Eq.~\eqref{eq:shearforces} into Eq.~\eqref{eq:p1} and Eq.~\eqref{eq:p2}, we get
\begin{align}
	c_{11} \left( r u_1'' + u_1' - \frac{u_1}{r} \right) + G \frac{u_2 - u_1}{d h} r &= -\omega^{2} \rho r u_1 \label{eq:p1b}, \\
	c_{11} \left( ru_2'' + u_2' - \frac{u_2}{r} \right) - 2 G \frac{u_2 - u_1}{d h} r &= -\omega^{2} \rho r u_2 \label{eq:p2b}.
\end{align}
We do not continue writing up the differential equation and boundary conditions of plate 3 since they are completely identical to those of plate 1.

Now introduce the variable
\begin{equation}
	u \equiv u_2 - u_1 = u_2 - (u_1 + u_3)/2.
\end{equation}
Subtracting Eq.~\eqref{eq:p1b} from Eq.~\eqref{eq:p2b}, we get
\begin{equation}\label{eq:p2p1}
	c_{11} \left( ru'' + u' - \frac{u}{r} \right) - 3 G \frac{u}{d h} r = -\omega^{2} \rho r u
\end{equation}
or
\begin{equation}\label{eq:p2p1b}
	r^{2} u'' + r u' + \left[ \left( \frac{\omega^{2} \rho}{c_{11}} - \frac{3 G}{c_{11} d h} \right) r^{2} - 1 \right] u = 0.
\end{equation}
Comparing to  Eq.~(20) in \citet{christensen_rheometer_1995}, we see that $d$ has been replaced by $d/3$.
However, we also have to see what happens to the boundary conditions and the expression for the normalized capacitance $F$.
The boundary conditions become zero displacement at the centres,
\begin{eqnarray}
	u_1(0) = 0, \quad u_2(0) = 0,
\end{eqnarray}	
and zero stress at the boundaries
\begin{equation}
	\begin{split}
		u_1'(R) + \frac{\nu}{R} u_1(R) = (1 + \nu) d_{13} E^{(1)}_z, \\
		u_2'(R) + \frac{\nu}{R} u_2(R) = (1 + \nu) d_{13} E^{(2)}_z.
	\end{split}
\end{equation}

The two outer discs are electrically coupled in series, and this couple is then connected in parallel with the middle disc.
Furthermore, the polarity of the middle disc is reversed compared to that of the outer discs to reverse the motion of the middle disc.
That means that
\begin{equation}
	E_z \equiv E^{(2)}_z = -2 E^{(1)}_z,
\end{equation}
and thus the boundary conditions for $u$ become
\begin{equation}
	u(0) = 0
\end{equation}
and
\begin{equation}
	u'(R) + \frac{\nu}{R} u(R) = (1 + \nu) d_{13} \frac{3}{2} E_z.
\end{equation}
If we normalize a little differently than for the one-one configuration \cite{christensen_rheometer_1995} and include the factor of $3/2$ in front of $E_z$,
\begin{equation}
	x \equiv \frac{r}{R}, \quad e(x) \equiv \frac{2}{3 (1 + \nu) d_{13} E_z R} u(R x),
\end{equation}
we get exactly the same differential equation and boundary conditions for $e(x)$ as in \citet{christensen_rheometer_1995},
\begin{align}
	x^{2} e'' + xe' + (k^{2} - 1)e &= 0, \\
	e(0) = 0, \quad e'(1) + \nu e(1) &= 1,
\end{align}
with $k$ given as in Eq.~\ref{eq:ksquare}.

This means that the solution in terms of $e(x)$ is exactly the same.
Only the scaling factor of the modulus $G_\text{c}$ is changed by $1/3$.
The normalized modulus $V = {G(\omega)}/{G_\text{c}}$ is the same for the 3-PZ PSG and the one-one configuration, but $G_\text{c}$ is changed.

Finally, let us discuss the measured capacitance $C_\text{m}$.
For each of the three discs, Eq.~(14) of \citet{christensen_rheometer_1995} is valid.
It can be written for disc 1 (and 3) as
\begin{align}
	\begin{split}
		C^{(1)}_m &= \left( C_\text{f} - C_\text{cl} \right) \frac{u_{1}(R)}{E^{(1)}_z R d_{13}} + C_\text{cl} \\
		&= \left( C_\text{f} - C_\text{cl} \right) \frac{-2 u_{1}(R)}{E_z R d_{13}} + C_\text{cl},
	\end{split}
\end{align}
and for disc 2 as
\begin{align}
	\begin{split}
		C^{(2)}_m &= \left( C_\text{f} - C_\text{cl} \right) \frac{u_2(R)}{E^{(2)}_z R d_{13}} + C_\text{cl} \\
		&= \left( C_\text{f} - C_\text{cl} \right) \frac{u_2(R)}{E_z R d_{13}} + C_\text{cl}.
	\end{split}
\end{align}
The total capacitance of the configuration becomes
\begin{align}
	\begin{split}
		C_\text{m} &= \frac{1}{2} C^{(1)}_m + C^{(2)}_m \\
		&= \left( C_\text{f} - C_\text{cl} \right) \frac{u_2(R) - u_1(R)}{E_z R d_{13}} + \frac{3}{2} C_\text{cl}\\
		&= \left( C_\text{f} - C_\text{cl} \right) \frac{3}{2} (1 + \nu) e(1) + \frac{3}{2} C_\text{cl}.
	\end{split}
\end{align}
Thus the normalized capacitance of the 3-PZ PSG,
\begin{equation}
	F = \frac{C_\text{m} - \frac{3}{2} C_\text{cl}}{\frac{3}{2} \left( C_\text{f} - C_\text{cl} \right)} = (1 + \nu) e(1),
\end{equation}
is exactly identical to that of the one-one configuration.
The only difference is that the effective liquid layer thickness is one third of the real layer in the 3-PZ PSG.

\section{Including effects of partial liquid filling}
\label{sec:including_effects_of_partial_liquid_filling_and_the_hub_in_the_psg}
The liquid sample is filled into the PSG at room temperature such that the sample diameter exactly matches the diameter of the discs.
But because of thermal contraction, when the PSG and sample are cooled down, the radius, $r$, of the liquid layer will be smaller than the disc radius, $R$.
This partial filling leads to the sample clamping the discs to a lesser extent.
Therefore, a lower modulus than the true value would be determined if the contraction was not taken into account in the inversion procedure.
In \citet{christensen_rheometer_1995}, an expression for the normalized capacitance, $F$, (defined by Eq.~\eqref{eq:Fdef}) was given, making $F = F(S,V,x_l)$ also a function of the scaled liquid radius, $x_l = r/R$.

First we review the analytic solution to the problem of partial filling including some more details on the derivations.
The differential equations shown as Eq.~(46) of \citet{christensen_rheometer_1995} were
\begin{align}
	\begin{split}
		x^2 e_1'' &+ x e_1' + \left[ (k_1 x)^2 - 1 \right] e_1 = 0, \\
		k_1^2 &= \left( \frac{\omega}{\omega_\text{c}} \right)^2 - \frac{G}{G_\text{c}}, \\
		x^2 e_2'' &+ xe_2' + \left[ (k_2 x)^2 - 1 \right] e_2 = 0, \\
		k_2^2 &= \left( \frac{\omega}{\omega_\text{c}} \right)^2,
	\end{split}
\end{align}
and the boundary conditions were
\begin{align}
	e_1(0) &= 0, \label{eq:bndcond1}\\
	e_1(x_l) &= e_2(x_l), \label{eq:bndcond2} \\
	e_1'(x_l) + \frac{\nu}{x_l} e_1(x_l) &= e_2'(x_l) + \frac{\nu}{x_l} e_2(x_l), \label{eq:bndcond3} \\
	e_2'(1) +\nu e_2(1) &= 1. \label{eq:bndcond4}
\end{align}
By using the second boundary condition, the third can be simplified to
\begin{equation}
	e_1'(x_l) = e_2'(x_l). \label{eq:bndcond3b}
\end{equation}
The solutions are, in terms of Bessel functions $J_1$ and $Y_1$,
\begin{eqnarray}
	e_1(x) = A J_1(k_1x) + B Y_1(k_1x) \label{eq:sol1}, \\
	e_2(x) = C J_1(k_2x) + D Y_1(k_2x) \label{eq:sol2}.
\end{eqnarray}

Now, $J_1(0) = 0$ and $Y_1(0) = -\infty$.
Thus Eq.~\eqref{eq:bndcond1} implies $B = 0$ and Eq.~\eqref{eq:bndcond2} becomes
\begin{equation}
	A J_1(k_1 x_l) = C J_1(k_2 x_l) + D Y_1(k_2 x_l). \label{eq:boundcond2b}
\end{equation}
The derivative of the Bessel function $ J_1 $ (and likewise $ Y_1 $) fulfils
\begin{equation}
	\dod{}{x} J_1(kx) = k J_0(kx) - \frac{1}{x} J_1(kx). \label{eq:Bessderiv}
\end{equation}
When inserting the solutions Eq.~\eqref{eq:sol1} and Eq.~\eqref{eq:sol2} into Eq.~\eqref{eq:bndcond3b} using Eq.~\eqref{eq:Bessderiv}, a number of terms with zeroth and first order Bessel functions appear.
The first order terms cancel due to Eq.~\eqref{eq:boundcond2b}, and we are left with
\begin{equation}
	A k_1 J_0(k_1 x_l) = C k_2 J_0(k_2 x_l) + D k_2 Y_0(k_2 x_l).
\end{equation}
Multiplying this equation with $J_1(k_1 x_l)$ and inserting Eq.~\eqref{eq:boundcond2b}, we get rid of the constant $A$,
\begin{align}
	\begin{split}
	C k_1 J_0(k_1 x_l) J_1(k_2 x_l) &+ D k_1 J_0(k_1 x_l) Y_1(k_2 x_l) = \\
	C k_2 J_0(k_2 x_l) J_1(k_1 x_l) &+ D k_2 Y_0(k_2 x_l) J_1(k_1 x_l),
	\end{split}
\end{align}
or, collecting terms,
\begin{equation}
	RC + TD = 0, \label{eq:bndcond3c}
\end{equation}
where
\begin{align}
	\begin{split}
	R &= k_1 x_l J_0(k_1 x_l) J_1(k_2 x_l) - k_2 x_l J_0(k_2 x_l) J_1(k_1 x_l), \\
	T &= k_1 x_l J_0(k_1 x_l) Y_1(k_2 x_l) - k_2 x_l Y_0(k_2 x_l) J_1(k_1 x_l).
	\end{split} \label{eq:RTdef}
\end{align}
(Note: The $ R $ and $ T $ terms could have been defined without the superfluous $ x_l $ factors outside the Bessel functions. However, we comply with the original definitions in \citet{christensen_rheometer_1995}).

Finally, the fourth boundary condition, Eq.~\eqref{eq:bndcond4}, yields
\begin{equation}
	PC + QD = 1, \label{eq:bndcond4b}
\end{equation}
where
\begin{align}
	\begin{split}
	P &= k_2 J_0(k_2) + (\nu - 1) J_1(k_2) \\
	Q &= k_2 Y_0(k_2) + (\nu - 1) Y_1(k_2),
	\end{split} \label{eq:PQdef}
\end{align}
once again using Eq.~\eqref{eq:Bessderiv}.
Define the determinant
\begin{equation}
	\Delta = PT - RQ,
\end{equation}
and the solution of Eq.~\eqref{eq:bndcond3c} and Eq.~\eqref{eq:bndcond4b} with respect to $C$ and $D$ is
\begin{eqnarray}
	C = T/\Delta, \quad D = -R/\Delta.
\end{eqnarray}

\section{Including the Inelastic Hub}
\label{sec:details_on_the_inelastic_hub_calculations}
Including an inelastic hub essentially just moves the first boundary condition to the relative radius $x_h = r_\text{h}/R$ of the hub.
Thus Eq.~\eqref{eq:bndcond1} is replaced by
\begin{equation}
	e_1(x_h) = 0.
\end{equation}
The solutions are still of the form Eq.~\eqref{eq:sol1} and Eq.~\eqref{eq:sol2}, but now $B \neq 0$.
For the sake of readability, let
\begin{align}
	a = J_1(k_1 x_h), &\quad b = Y_1(k_1 x_h), \\
	c = J_1(k_1 x_l), &\quad d = Y_1(k_1 x_l), \\
	e = J_1(k_2 x_l), &\quad f = Y_1(k_2 x_l), \\
	g = k_1 x_l J_0(k_1 x_l), &\quad h = k_1 x_l Y_0(k_1 x_l), \\
	i = k_2 x_l J_0(k_2 x_l), &\quad j = k_2 x_l Y_0(k_2 x_l),
\end{align}
and the three first boundary conditions become
\begin{align}
	Aa + Bb &= 0, \label{eq:bndcond1d} \\
	Ac + Bd &= Ce + Df, \label{eq:bndcond2d} \\
	Ag + Bh &= Ci + Dj\,, \label{eq:bndcond3d}
\end{align}
while the fourth is still
\begin{equation}
	PC + QD = 1.
\end{equation}
The last equation is the unaltered Eq.~\eqref{eq:bndcond4b} with $P$ and $Q$ still defined by Eq.~\eqref{eq:PQdef}.

Multiplying Eq.~\eqref{eq:bndcond2d} and Eq.~\eqref{eq:bndcond3d} with $b$ and using Eq.~\eqref{eq:bndcond1d} to eliminate $B$, the second and third boundary conditions become
\begin{align}
	A(b c - a d) &= C b e + D b f, \label{eq:bndcond2e} \\
	A(b g - a h) &= C b i + D b j. \label{eq:bndcond3e}
\end{align}
Multiplying Eq.~\eqref{eq:bndcond3e} with $bc - ad$ and using Eq.~\eqref{eq:bndcond2e}, the third boundary condition becomes
\begin{equation}
	(C b e + D b f)(b g - a h) = (C b i + D b j)(b c - a d)
\end{equation}
or (cancelling a common $b$ factor and collecting terms)
\begin{align}
	\begin{split}
		&\left[ b(g e - c i) + a(d i - h e) \right] C \\
		&\quad + \left[ b(f g - c j) + a(d j - f h) \right] D = 0.
	\end{split}
\end{align}
We recognise that
\begin{align}
	R = g e - c i, \quad T = f g - c j
\end{align}
as defined in Eq.~\eqref{eq:RTdef} for the problem without a hub.
Define
\begin{align}
	U = d i - h e, \quad W = d j - f h
\end{align}
and
\begin{align}
	R' = b R + a U, \quad T' = b T + a W.
\end{align}
Then the third boundary condition of the problem with the hub becomes
\begin{equation}
	R' C + T' D = 0, \label{eq:bndcond3f}
\end{equation}
which resembles that of the problem without hub.
The solution likewise becomes
\begin{align}
	C = T'/\Delta', \quad D = -R'/\Delta' \label{eq:CDsol}
\end{align}
with
\begin{equation}
	\Delta' = P T' - R' Q.
\end{equation}

Written in this form, it is easy to see that the solutions for the two problems coincide when $x_h \rightarrow 0$.
Then $a \rightarrow 0$, since $J_1(0) = 0$.
On the other hand, $b \rightarrow -\infty$.
However, since $R' \rightarrow b R$ and $T' \rightarrow b T$, it follows that  $\Delta' \rightarrow b \Delta$.
Thus the common diverging $b$ factors of the denominator and the numerator in Eq.~\eqref{eq:CDsol} cancel, giving the right limiting behaviour.

Now $V = G(\omega)/G_\text{c}$ can be found by inversion of
\begin{eqnarray}
	F(S,V,x_l) = (1 + \nu) \left( C J_1(k_2) + D Y_1(k_2) \right), \label{eq:Fshear_rh} 
\end{eqnarray}
recalling that the left-hand side can be measured and $C$ and $D$ depend on $V$  via $k_1$.
When including the hub, we can still determine $\omega_\text{c}$ by a fit to the resonance spectrum of the empty transducer, but the relation to the first resonance is changed.
Now $\omega_1 \neq 2.054 \omega_\text{c}$.

\section{Measurements on a solid sample (Rubber)}
\label{sec:rubber}
\begin{figure}
	\includegraphics{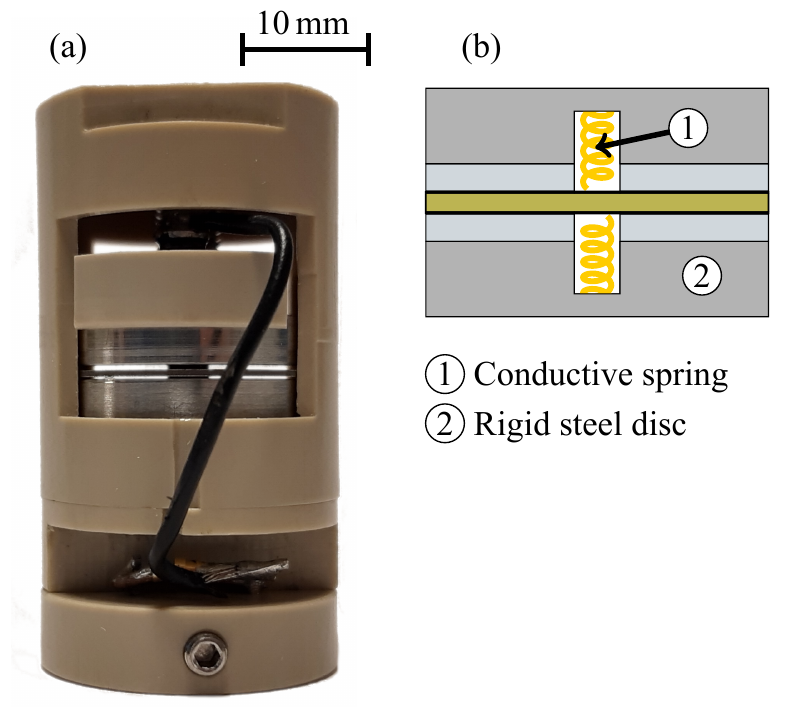}
	\caption{
	a) Photo of 1-PZ PSG for solid samples.
	The assembly of PZ disc and rigid supports is encased in a plastic housing, which may be opened to take the discs apart for sample mounting.
	The rigid supports are made from steel and form part of the electrical connection to the PZ disc.
	b) Schematic drawing of the centre of the device.
	Metal springs placed in recesses in the steel plates make electrical connection to the PZ disc, when a sample is mounted.
	}
	\label{fig:PSG_Steel}
\end{figure}
Figure \ref{fig:PSG_Steel}(a) shows a photo of the 1-PZ PSG implementations for solid samples. Here the rigid supports are steel discs and the assembly is encased in a PEEK housing. This assembly can easily be taken apart so that solid samples can be mounted in the cell. A solid sample needs to be glued to the supports as well as the PZ disc to ensure a no-slip boundary condition in the measurement. 
The centering of the disc, sample and rigid supports is ensured in the gluing process: the “sample sandwich” (outer plate – sample disc – PZ disc – sample disc – outer plate, see Fig.~\ref{fig:PSG_Steel}(b)) is glued inside a Teflon mold that ensures all layers are perfectly aligned.
Parallelism is ensured if all layers have a uniform thickness.

A schematic drawing of the inner part is shown in Fig.~\ref{fig:PSG_Steel}(b). The electrical connection to the PZ disc is secured through the steel supports by small springs in the centre going through a small hole in the sample to make contact with the electrodes on the PZ disc. In this case, the sample itself acts as a spacer between the steel support and the PZ disc.

\begin{figure*}
	\includegraphics{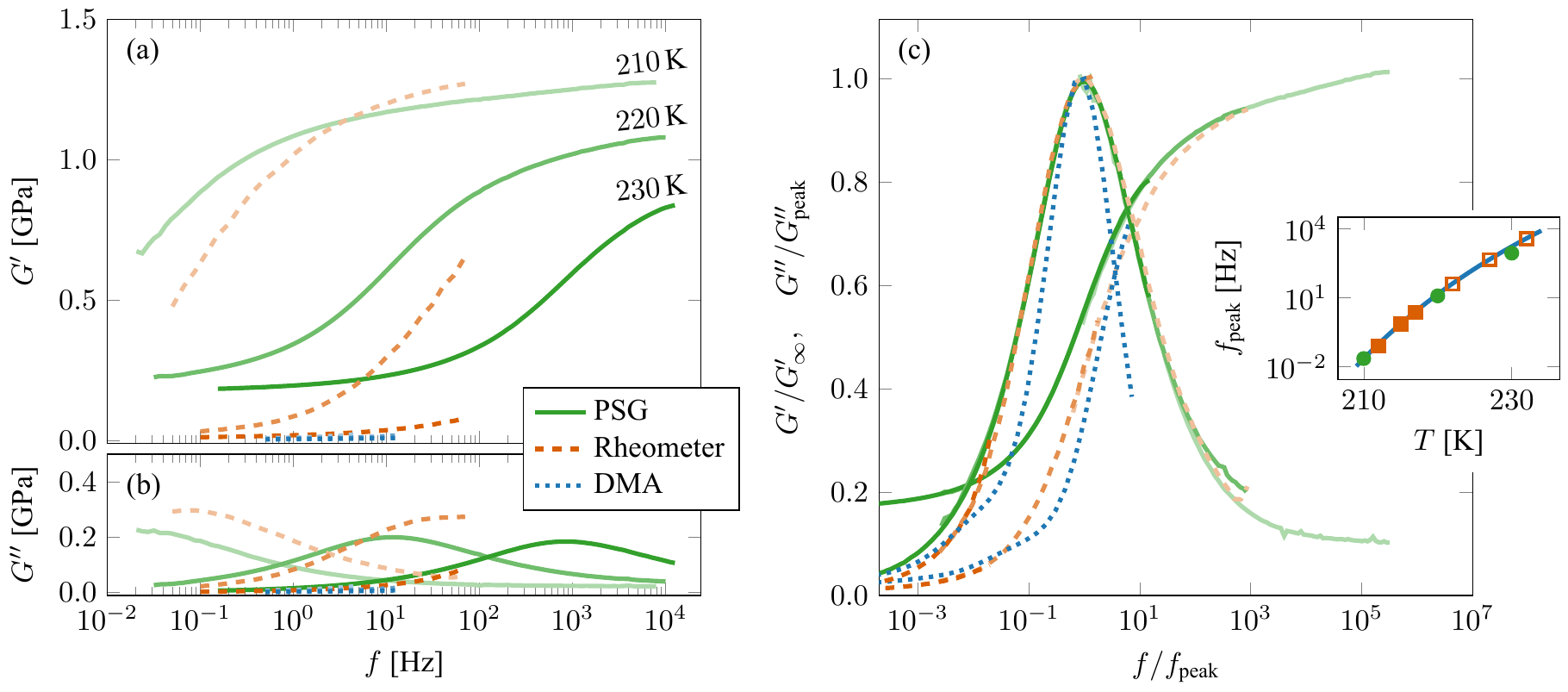}
	\caption{
		Shear modulus of filled rubber, measured both by PSG, standard rheometer and dynamic mechanical analysis (DMA).
		Solid lines are measurements made by PSG at \SIlist{210;220;230}{\kelvin}.
		Dashed lines are standard rheometer measurements made at the same temperatures.
		Dotted lines are DMA measurements carried out at \SIlist{223;228;233}{\kelvin}.
		a) and b) show storage and loss moduli, while c) shows both, scaled by the high-frequency limit of $G'$, denoted $G_\infty$, and the value of $G''$ at the loss peak, $G''_{\text{peak}}$, respectively.
		The inset in c) shows the temperature dependence of the loss-peak position for the three measurement series.
		For the PSG measurements (filled circles), as well as the lowest-temperature rheometer measurements (filled squares), the peak positions were determined directly from the loss modulus spectra.
		For higher-temperature rheometer measurements (unfilled squares) and all DMA measurements (full line) the positions were found by mastering and shifted to match the value of the PSG measurement at \SI{210}{\kelvin} (see main text for more details).
	}
	\label{fig:RubberModulus}
\end{figure*}

As a proof of concept, measurements on a rubber sample are shown in Fig.~\ref{fig:RubberModulus}.
The sample used is a synthetic isoprene rubber (IR) with carbon black N121 filler.
The recipe (see Tab.~\ref{tab:rubber_recipe}) was defined in analogy to ASTM~D3191\cite{astm_international_astm_2021}, but using synthetic polyisoprene instead of styrene-butadiene rubber and adding anti-aging additives (wax, TMQ: oligomers of 1,2-dihydro-2,2,4-trimethylquinoline
6PPD: N-(1,3-dimethylbutyl)-N'-phenyl-p-phenylenediamine).
The specification of N121 can be found in ASTM~D1765\cite{astm_international_astm_2021-1}.

\begin{table}
	\centering
	\begin{tabular}{lS}
		\toprule
		Substance & {Amount [phr]} \\
		\midrule
		Polyisoprene\footnotemark[1] & 100 \\
		N121\footnotemark[2] & 50 \\
		Additives\footnotemark[3] & 6 \\
		Zink oxide\footnotemark[4] & 3 \\
		Stearic acid\footnotemark[4] & 2 \\
		Sulfur\footnotemark[5] & 1.4 \\
		TBBS\footnotemark[5] & 0.75 \\
		\bottomrule
		\noalign{\vskip 1mm} 
		\multicolumn{2}{l}{\footnotemark[1]polymer} \\
		\multicolumn{2}{l}{\footnotemark[2]filler} \\
		\multicolumn{2}{l}{\footnotemark[3]e.g. anti-aging} \\
		\multicolumn{2}{l}{\footnotemark[4]vulcanization activator} \\
		\multicolumn{2}{l}{\footnotemark[5]vulcanization package}
	\end{tabular}
	\caption{
		Components of the IR+N121 rubber compound.
		All amounts are given by weight relative to 100 units of weight of the rubber polymer (\emph{per hundred rubber} or phr).
		TBBS is N-tert-butyl-2-benzothiazolesulfenamide.
	}
	\label{tab:rubber_recipe}
\end{table}

The shear modulus was measured by the 1-PZ PSG at temperatures from \SIrange{210}{230}{\kelvin} in steps of \SI{10}{\kelvin}.
At these temperatures, the shear loss peak is in the frequency window of the PSG.
At the peaks, the frequency, $f_\text{peak}$, and loss modulus, $G''_\text{peak}$, were determined, and along with the high-frequency storage modulus, $G_\infty$, these values were used to construct the normalized master curve, shown in Fig.~\ref{fig:RubberModulus}(c).

For comparison, the shear modulus was also measured with a commercial rheometer (Modular Compact Rheometer MCR 502, Anton Paar, Graz, Austria) as well as by dynamic mechanical analysis (DMA Gabo Eplexor$^\circledR$ 2000N, Netzsch, Alhden, Germany).

The rheometer measurements were conducted in oscillatory mode with a deformation of \SI{0.02}{\percent} in a temperature range from \SIrange{210}{230}{\kelvin}.
A frequency sweep was performed (from \SIrange{0,1}{70}{\hertz}) at each temperature and the complex shear modulus was recorded.
For the lowest temperatures, the loss peak frequency and modulus could be determined, and this was used as the basis for a master curve, along with shift factors determined by matching $\tan \delta$ at higher temperatures.

The DMA measurements were conducted in compression with a deformation of up to \SI{2}{\percent} in a temperature range from \SIrange{-55}{80}{\celsius} (\SIrange{218}{353}{\kelvin}).
A frequency sweep was performed (from \SIrange{0.5}{50}{\hertz}) every \SI{5}{\kelvin} and the complex Young's modulus, $E$, was determined.
Young's modulus was converted\cite{landau_theory_1986} into shear modulus using $G = E/(2 (1 + \nu))$.
For rubber where the Poisson ratio is $\nu = 1/2$ this becomes $G = (1/3) E$.
A master curve was then produced using Williams-Landel-Ferry shift factors\cite{williams_temperature_1955}.
From these shift factors, we also obtain the temperature dependence of the material times.

The frequency window of the DMA measurement is rather narrow and measurements were only carried out in the soft region around \SIrange{1}{20}{\mega\pascal}, whereas the PSG and rheometer measurements resolve moduli from around \SI{1}{\mega\pascal} up to several \SI{}{\giga\pascal}.

Figure \ref{fig:RubberModulus}(a+b) shows real and imaginary part of all three measurements and Fig.~\ref{fig:RubberModulus}(c) shows the corresponding mastercurves.
The real part of PSG and rheometer measurements agree within \SI{10}{\percent} on a high-frequency plateau, $G_\infty$, of about \SI{1.2}{\giga\pascal} (Fig.~\ref{fig:RubberModulus}(a)), while the spectral shape of the imaginary parts are nearly identical (Fig.~\ref{fig:RubberModulus}(c)).
DMA and rheometer measurements agree in the low-frequency region in their estimation of the rubber plateau.
In this region, the PSG measurements gives a value that is 20 times higher.
This could be due to the relatively low sensitivity at low moduli of the PSG measurement or it could be an artifact introduced by for instance the glue.
Work is ongoing to pinpoint the origin of this discrepancy.

The inset of Fig.~\ref{fig:RubberModulus}(c) shows the temperature dependence of the characteristic time scales for relaxation derived from the three measurements.
Using the PSG loss peak frequencies (full circles) as a reference, time scales from rheometer (squares) and DMA (full line) measurements are shifted to ensure agreement with the PSG loss peak frequency at \SI{210}{\kelvin}.
The shift for the rheometer measurement was \SI{+2}{\kelvin} on the temperature axis and reflects a difference in absolute temperature calibration between the setups. 
For the DMA measurements the Williams-Landel-Ferry curve for the shift factors was converted into frequencies and shifted on the $y$-axis to match the PSG measurements at \SI{210}{\kelvin}.
Note that only for the PSG measurements can the loss peaks be directly determined in the full temperature range.
Nevertheless, the temperature dependence of the time scales for all three measurements match remarkably well.

\section{Including liquid inertia.}  \label{sec:Appinerti}
In the derivation of the equation of motion of the piezoelectric disc, it was assumed that the inertia of the liquid could be ignored in the liquid stress response to the disc displacement.
In order to take the inertia into account we consider the wave-equation for shear waves in the $z$-direction
\begin{equation}
	 -\omega^2 u_x(z) = \frac{G(\omega)}{\rho_l}\dod[2]{u_x(z)}{z},
\end{equation}
with boundary conditions
\begin{equation}
	u_z(0) = 0, \quad \sigma_l = G(\omega) \left( \dod{u_x(z)}{z}\right)_{z = d}\,.
\end{equation}
Note that $G(\omega)$ (here, as well as in the main manuscript) denotes the complex modulus, $G = \Gprime + i\Gdoubleprime$.

\begin{figure}
	\centering
	\includegraphics{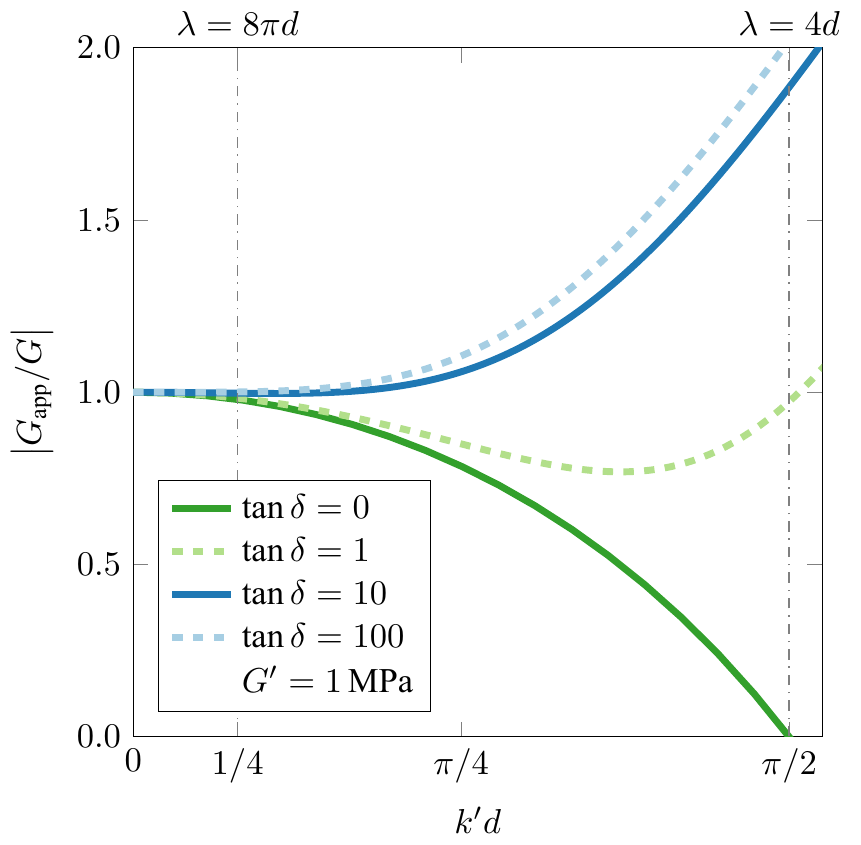}
    \caption{
		$\left| G_\text{app}/{G} \right|$ as a function of $k^\prime d$ for the real modulus of $\Gprime = \SI{1}{\giga\pascal}$ for four values of $\tan \delta$: 0 (the elastic case), 1, 10 and 100 (the very viscous case).
		Two values of $\lambda = 2\pi/k^\prime$ are highlighted:
		$\lambda = 8 \pi d$ ($k^\prime d = 1/4$) where, for a purely real $G$, $\left| G_\text{app}/{G} \right| = \num{0.98}$ and $\lambda = 4 d$ ($k^\prime d = \pi/2$) where, for a purely real $G$, $\left| G_\text{app}/{G} \right| = \num{0}$ and the correction factor diverges.
		For a finite imaginary part, the correction factor is always finite (i.e.~the curves for $\tan\delta\neq0$ never vanish), meaning that a correction is always possible.
		In the region between the two vertical lines, a correction for sample inertia is necessary.
		For $\tan\delta\neq0$, it is in principle possible to correct also above $k'd = \pi/2$, but the PSG method is not meaningful when the transverse sound wavelength becomes comparable to the sample thickness.
		This thus marks the lower limit for the method for all moduli.
		See also Fig.~\ref{fig:Inertial_Limit.
		}
	}
	\label{fig:inertial_correction}
\end{figure}

Solving these equations one finds that the modulus determined by neglecting inertial effects -- the apparent modulus $G_\text{app}(\omega)$ -- is a function of actual modulus $G(\omega)$,
\begin{equation} \label{eq:a3}
	G_\text{app}(\omega) \equiv d \frac{\sigma_l}{u_x(d)} = G(\omega) \frac{kd}{\tan(kd)} ,
\end{equation}
with
\begin{equation}
	k = \sqrt{\frac{\rho_l}{G(\omega)}}\omega\,,
\end{equation}
where $\rho_l$ is the sample density.
Note that in the general case where the modulus $G$ is complex, $k$ is also complex, $k = k^{\prime} + i k^{\prime\prime}$, with $k^{\prime}$ representing the wavenumber of the transverse sound wave and $k^{\prime\prime}$ the damping coefficient.

In order to correct for the effect of liquid inertia, one has to invert $G_\text{app}$ (as determined by the method described in Secs.~\ref{sec:the_piezoelectric_shear_modulus_gauge} and~\ref{sec:inversion_algorithm}) by Eq.~\eqref{eq:a3}.
This is done by introducing
\begin{equation}
    a = \frac{\rho_l \omega^2 d^2}{G_\text{app}}\quad \text{ and } x = kd,
\end{equation}
whereby Eq.~\eqref{eq:a3} becomes 
\begin{equation}
    a = x\tan(x),
\end{equation}
which is solved for $x$ from which $G(\omega)$ is found.

Figure~\ref{fig:inertial_correction} shows the ratio of $\left| G_\text{app}/G\right|$ as a function of $k'd$for a $G' = 1$~MPa and a range of $\tan\delta$ from 0 (the purely elastic case) to 100 (a very viscous case). The plot illustrates that the correction factor starts to matter when $k'd \approx 1/4$.
Two values for $k'd$ are marked by dashed lines:
1) where the correction becomes sizeable (corresponding to a transverse sound wavelength of $\lambda = 8 \pi d$) and
2) the theoretical limit for the correction in the purely elastic case, where $|G_\text{app}/G| = 0$. This corresponds to a transverse sound wavelength of $\lambda = 4d$.
When the modulus is complex there is no theoretical limit to the correction; $|G_\text{app}/G|$ never vanishes and thus the correction factor never diverges.
However, we still consider $\lambda = 4d$ (equivalent to $k'd = \pi/2$ in the purely elastic case) to be the practical limit for the PSG method.
\begin{figure}
	\centering
	\includegraphics{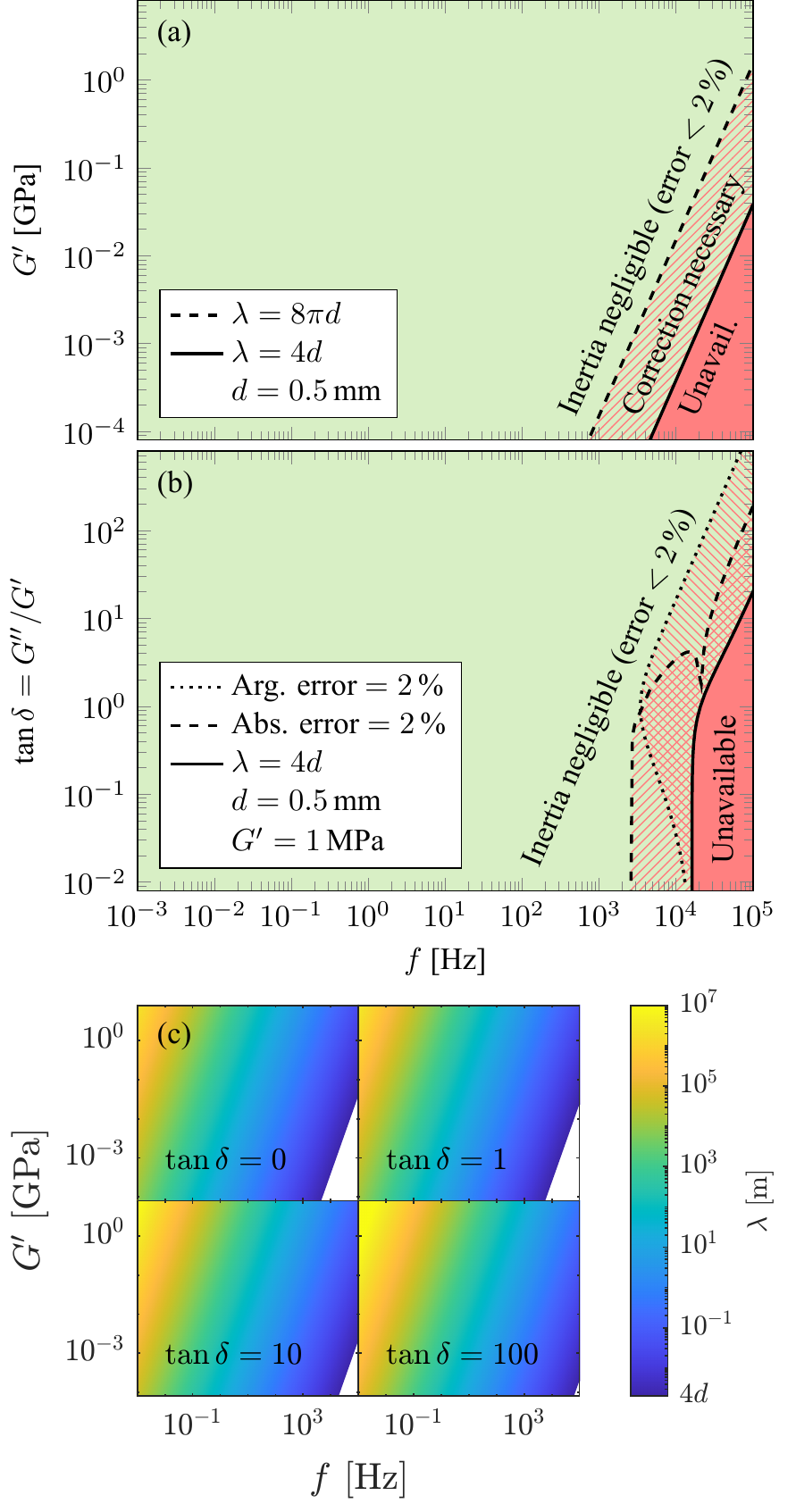}
	\caption{
		Illustration of the necessity of correcting for inertial effects for a sample thickness $d = \SI{0.5}{\milli\metre}$.
		a) The elastic case, i.e.~$\Gdoubleprime = 0$.
		In most of the plane, inertial effects are negligible, but at a combination of high frequency and low modulus, the wavelength of the shear wave in the sample becomes comparable to the sample thickness.
		For $\lambda = 8 \pi d$ (shown as a dashed line), we have  $G_\text{app}/G \approx 0.98$.
		Hence, the correction is less than \SI{2}{\percent} for the area to the left of this line.
		The solid line, where $\lambda = 4 d$, marks the limit of the correction where $G_\text{app}/G = 0$ and the correction factor diverges.
		b) Like a), but showing the situation for a fixed $\Gprime = \SI{1}{\mega\pascal}$ when varying $\tan\delta$.
		In this case, the dashed line marks a 2~\% error in the absolute correction factor $|G_\text{app}/G|$ and the dotted line marks the 2~\% error in the phase, $\arg(G_\text{app}/G)$.
		Inertia is negligible in the area to the left of the union of these two areas.
		c) The wavelength of the shear wave for four values of $\tan\delta$.
		The areas where $\lambda < 4d$ are shown in white.
		Note how the white area shrinks for higher $\tan\delta$.
	}
	\label{fig:Inertial_Limit}
\end{figure}

Figure~\ref{fig:Inertial_Limit} shows where the practical limit of the method lies in the frequency-vs.-modulus plane.
The limit is determined by $\lambda = 4d$ as well as where the effect of inertia becomes significant.
The elastic case ($G'' = 0$) is shown in Fig.~\ref{fig:Inertial_Limit}(a).
The limits on the frequency and modulus axes correspond to the range of the PSG measurement.
The red area in the lower right corner marks the practical limit of the PSG measurement, i.e.~where $\lambda = 4d$, below which the method cannot measure meaningfully.
The dashed line marks where the inertia correction is 2~\% and the hatched area thus shows where correction is necessary.
The visco-elastic case is illustrated in Fig.~\ref{fig:Inertial_Limit}(b) in a plot similar to Fig.~\ref{fig:Inertial_Limit}(a), except here $\tan\delta$ is varied (equivalent to adding an imaginary part) for a fixed $G' = 1$~MPa.
In this plot, the practical limit of the measurement is again the solid line in the lower right corner.
The dashed and dotted lines mark where the absolute value, respectively the phase, of the correction factor $G_\text{app}/G$ are 2~\%.
In Fig.~\ref{fig:Inertial_Limit}(c) $G'$ is varied for different fixed values of $\tan\delta$.
The colour shows the wavelength of the transverse sound wave and the white area where $\lambda \leq 4d$.
Clearly, this area shrinks as the imaginary part grows.
Note that the locations of ``forbidden'' areas depend on the sample thickness.
All illustrations used here are for $d = 0.5$~mm.
The limit of the method ($\lambda \leq 4d$) goes as $\propto d^2$, so a smaller sample thickness would lower the limit.

\begin{figure}
	\centering
	\includegraphics{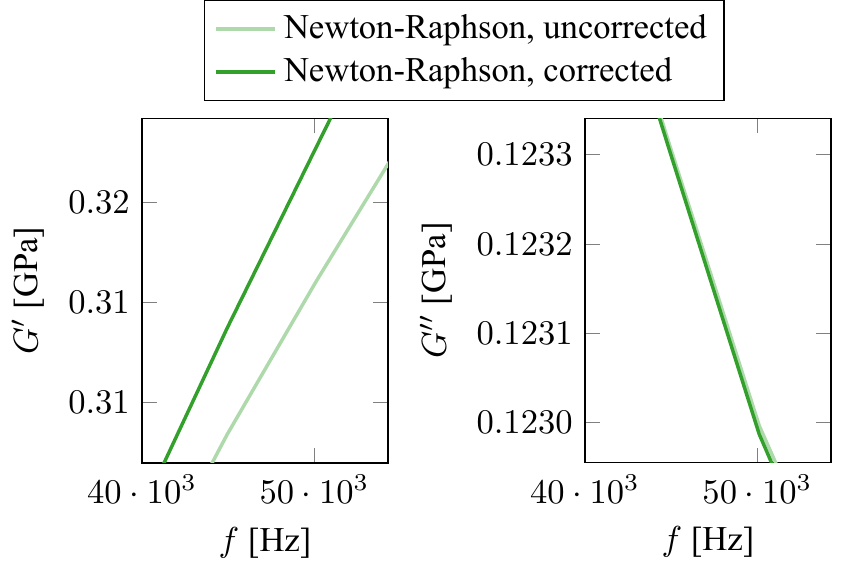}
	\caption{
	    Comparison of uncorrected and corrected data, $G_\text{app}$ and $G$, for squalane measured at \SI{190}{\kelvin}.
	    These data are also shown in fig.~\ref{fig:Inversion_Comparison}.
	    The correction on the loss modulus is relatively much smaller than that on the storage modulus.
	    Note how the correction raises the storage modulus from the measured value, while it lowers the loss modulus.}
	\label{fig:Inversion_Comparison_Zoom}
\end{figure}

Finally, in Fig.~\ref{fig:Inversion_Comparison_Zoom} we show a zoom on the high frequency region of the real and imaginary part of uncorrected ($G_\text{app}'$ and $G_\text{app}''$ in light green) and corrected ($G'$ and $G''$ in dark green) data for squalane (same data as in Fig.~\ref{fig:Inversion_Comparison}) to demonstrate the effect on our data.
The corrected real part is slightly higher than the uncorrected, while the correction shifts the imaginary part to a very slightly lower value.

The correction for inertial effects derived here is identical to that of \citeauthor{schrag_deviation_1977}\cite{schrag_deviation_1977}.
The only difference is that we keep the complex notation instead of splitting the calculations in absolute modulus and phase, which makes the equations simpler and more operational for our experimental realisation.

\section{Plate bending calculations}\label{sec:platebending}
The Kelvin functions are defined as
\begin{align}
	\begin{split}
		\mathrm{Ber}_n(r) + i \mathrm{Bei}_n(r) &= J_n\left( r e^{3 \pi i/4} \right) \\
		\mathrm{Ker}_n(r) + i \mathrm{Kei}_n(r) &= e^{-\pi n i/2} K_n\left( r e^{\pi i/4} \right)
	\end{split}
\end{align}
where $J_n$ are the Bessel functions and $K_n$ are the modified Bessel functions of the first kind.
The following recursion formulas hold for the derivatives
\begin{align}
	\begin{split}
		\mathrm{Ber}' &= \frac{1}{\sqrt{2}} \left( \mathrm{Ber}_1 + \mathrm{Bei}_1 \right), \\
		\mathrm{Bei}' &= \frac{1}{\sqrt{2}} \left( \mathrm{Bei}_1 - \mathrm{Ber}_1 \right), \\
		\mathrm{Ber}_1' &=- \frac{1}{\sqrt{2}} \left( \mathrm{Ber} + \mathrm{Bei} \right) - \frac{1}{r}\mathrm{Ber}_1, \\
		\mathrm{Bei}_1' &=- \frac{1}{\sqrt{2}} \left( \mathrm{Bei} - \mathrm{Ber} \right) - \frac{1}{r}\mathrm{Bei}_1,
	\end{split}
\end{align}
and identical relations for the Ke-functions.

Define the Laplacian $\laplacian = \frac{1}{r} \pd{}{r} r \pd{}{r}$.
The recursion formulas lead to
\begin{align}
	\begin{split}
		\laplacian \mathrm{Ber} &= -\mathrm{Bei}, \\
		\laplacian \mathrm{Bei} &= \mathrm{Ber}, \\
		\laplacian \mathrm{Ker} &= -\mathrm{Kei}, \\
		\laplacian \mathrm{Kei} &= \mathrm{Ker},
	\end{split}
\end{align}
and thus all four Kelvin functions fulfil
\begin{equation}
		\laplacian^2 u = -u.
\end{equation}

The solution to Eq.~\eqref{eq:platedif3} of the main text is Eq.~\eqref{eq:bendingsolution}, which we repeat here:
\begin{equation}
	u(r) = A_1 \mathrm{Ber} (r) + A_2 \mathrm{Bei}(r) + A_3 \mathrm{Ker}(r) + A_4 \mathrm{Kei}(r)\,.
\end{equation}

The four boundary conditions of Eqs.~\eqref{eq:bnd4} and \eqref{eq:bnd1}-\eqref{eq:bnd3}  then become
\begin{align}
	\begin{split}
	u_0 &= A_1 \mathrm{Ber}(r_\text{h}) + A_2 \mathrm{Bei}(r_\text{h}) \\
					&+ A_3 \mathrm{Ker}(r_\text{h}) + A_4 \mathrm{Kei}(r_\text{h}),
	\end{split}
\end{align}
\begin{align}
	\begin{split}
		0 &= A_1 \left( \mathrm{Ber_1}(r_\text{h}) + \mathrm{Bei_1}(r_\text{h}) \right) \\
			&+ A_2 \left( \mathrm{Bei_1}(r_\text{h}) - \mathrm{Ber_1}(r_\text{h}) \right) \\
			&+ A_3 \left( \mathrm{Ker_1}(r_\text{h}) + \mathrm{Kei_1}(r_\text{h}) \right)\\
			&+ A_4 \left( \mathrm{Kei_1}(r_\text{h}) - \mathrm{Ker_1}(r_\text{h}) \right),
	\end{split}
\end{align}
\begin{align}
	\begin{split}
		0 &= A_1 \left( \mathrm{Ber_1}(R) - \mathrm{Bei_1}(R) \right) \\
			&+ A_2 \left( \mathrm{Ber_1}(R) + \mathrm{Bei_1}(R) \right) \\
			&+ A_3 \left( \mathrm{Ker_1}(R) - \mathrm{Kei_1}(R) \right) \\
			&+ A_4 \left( \mathrm{Ker_1}(R) + \mathrm{Kei_1}(R) \right)
	\end{split}
\end{align}
and
\begin{align}
	\begin{split}
		0 &= A_1 \left[ -\mathrm{Bei}(R) - g \left( \mathrm{Ber_1}(R) + \mathrm{Bei_1} (R) \right) \right] \\
			&+ A_2 \left[ \phantom{-} \mathrm{Ber}(R) - g \left( \mathrm{Bei_1}(R) - \mathrm{Ber_1}(R) \right) \right] \\
			&+ A_3 \left[ -\mathrm{Kei}(R) - g \left( \mathrm{Ker_1}(R) + \mathrm{Kei_1}(R) \right) \right] \\
			&+ A_2 \left[ \phantom{-} \mathrm{Ker}(R) - g \left( \mathrm{Kei_1}(R) - \mathrm{Ker_1}(R) \right) \right],
	\end{split}
\end{align}

where $g = (1 - \nu)/(\sqrt{2} R)$.

\end{document}